\documentclass[preprint,preprintnumbers,a4paper,11pt]{article}

\pdfoutput=1 

\usepackage{jheppub} 
\usepackage[T1]{fontenc} 
\usepackage{tikz}  
\usetikzlibrary{arrows.meta,positioning,calc,decorations.markings,shapes.misc}
\usepackage{tkz-graph}
\usepackage{array,amsfonts,mathtools}
\usepackage{enumerate}
\usepackage{adjustbox}
\usepackage{gensymb}
\usepackage{amsmath}
\usepackage{etoolbox}
\makeatletter

\begin{document}

\preprint{FTPI-MINN-24-09,\,\,\, UMN-TH-4318/24}

\title{New Soft Theorems for Two-Scalar Sigma Models}
\author[a]{Karol Kampf,}
\author[a]{Jiri Novotny,}
\author[b]{Mikhail Shifman,}
\author[c,a]{Jaroslav Trnka}

\affiliation[a]{Institute of Particle and Nuclear Physics, Faculty of
Mathematics and Physics, \\ Charles University, V Hole\v{s}ovi\v{c}k\'{a}ch 2, CZ-18000 Prague, Czech
Republic}
\affiliation[b]{William I. Fine Theoretical Physics Institute,\\ University of Minnesota,  Minneapolis, MN, 55455, USA}
\affiliation[c]{Center for Quantum Mathematics and Physics (QMAP), \\ University of California, Davis, California, USA}

\abstract{In this paper, we study the scattering amplitudes and soft theorems for the sigma models with two scalars. We show that if the particles are Goldstone bosons, then you necessarily get Adler zero with no possibility for non-trivial soft theorems. For non-Goldstone bosons, the soft behavior is generically captured by the geometric soft theorem studied by Cheung et al., and the right-hand side contains derivatives of lower-point amplitudes. Inspired by the recent work on the 2D sigma models, we study one special two-scalar sigma model, where the presence of symmetries in the target space translates into a special but non-trivial soft theorem without derivatives. We further generalize the construction to two larger classes of such models and derive certain soft theorem sum rules, again avoiding the derivatives of amplitudes. Our analysis provides an interesting hierarchy of two-scalar sigma models and soft theorems, ranging from Goldstone boson case to a generic target space, and showing that there are interesting theories in between.}

\maketitle

\section{Introduction and overview of the results}

The study of soft limits has always played an important role in our understanding of scattering amplitudes and the relation to symmetries of the underlying Lagrangian. In seminal paper \cite{Adler:1964um}, it was shown that the amplitudes in the $SU(N)$ non-linear sigma model vanish in the soft limit where one of the momenta $p\rightarrow0$. The corresponding sigma model describes the scattering of the Goldstone bosons. Relatively recently, the interest in amplitude soft limits was
revived in connection with the general resurrection of the $S$ matrix approach, where the kinematical properties, such as the soft limit, take the central role and can be used to define a theory.

The notion of the Adler zero was generalized in various directions. In particular, it was realized that in the soft limit, when the momentum $p$ of an external particle becomes soft, the amplitude may exhibit a stronger vanishing than a simple Adler zero, $A(p)\sim p^{\sigma }$ where a non-negative integer $\sigma >1$ \cite{Cheung:2014dqa}. It turned out that a given pattern of the soft limits of tree scalar-particle amplitudes combined with the Lorentz invariance allows one to unambiguously reconstruct underlying Lagrangians (assuming that they are local) \cite{Cheung:2014dqa}. In other words, the conventional strategy was reversed -- from the soft limit patterns to Lagrangians, rather than from Lagrangians to the generalized Adler zeros. A systematic classification was carried out in \cite{Cheung:2016drk}. Non-trivial soft behavior was proved to include the following important ``exceptional'' models whose interactions are fully determined by a single coupling constant: the $SU(N)$ non-linear sigma models (NLSM) \cite{Gell-Mann:1960mvl, Weinberg:1966fm}), 
NLSM with Wess-Zumino-Novikov-Witten terms \cite{Wess:1971yu, Novikov, Witten:1983tw}, Dirac-Born-Infeld (DBI) theory \cite{Born:1934gh} and Galileons \cite{Horndeski:1974wa, Nicolis:2008in} -- here the soft limit specified a special example of Galileon theories, known now as a special Galileon \cite{Cheung:2014dqa,Hinterbichler:2015pqa,Cachazo:2014xea}.  Amplitudes in these special theories can be then reconstructed using on-shell recursion relations \cite{Cheung:2015ota} which is a generalization of the BCFW prescription \cite{Britto:2004ap,Britto:2005fq} accounting for the presence of the soft limit behavior. The same set of theories also appears in the Cachazo-He-Yuan (CHY) formula \cite{Cachazo:2013hca,Cachazo:2013iea} and color-kinematics duality \cite{Bern:2008qj,Bern:2010ue,Bern:2010yg}, and form a set of exceptional theories, in some sense the simplest scalar effective field theories. 
The studies are not limited to scalar particles and tree-level amplitudes. The other spins were for example considered in this context in \cite{Cheung:2018oki,Elvang:2018dco,Kampf:2021bet,Kampf:2021tbk} and the first investigations of the NLSM integrand level in \cite{Bartsch:2022pyi,Bartsch:2024ofb}.
Very recently, the NLSM also played an important role in the surfacehedron construction of scattering amplitudes \cite{Arkani-Hamed:2023lbd,Arkani-Hamed:2023mvg}. In fact, the NLSM amplitudes can be obtained from a certain shift of the ${\rm Tr}\phi^3$ amplitudes \cite{Arkani-Hamed:2024nhp,Arkani-Hamed:2024yvu}, and satisfy further interesting properties such as hidden zeroes \cite{Arkani-Hamed:2023swr} and splits \cite{Arkani-Hamed:2024fyd} which seem to further extend to other exceptional theories \cite{Bartsch:2024amu,Li:2024qfp,Cao:2024gln}.

There is an interesting generalization of the Adler zero to cases where the amplitude has a non-zero soft limit, but the soft limit can be calculated and corresponds to a lower-point amplitude. For amplitudes containing dilatons of spontaneously-broken conformal symmetry it was discussed in \cite{Luo:2015tat}. Recently, it was also studied in \cite{Kampf:2019mcd} in the context of the $CP^1$ fibered model. This is a theory of Goldstone bosons which violates certain assumptions needed for the Adler zero, and hence the right-hand side of the soft limit is non-zero but can be expressed as a linear combination of lower-point amplitudes in the same theory. A different -- geometric -- perspective on the soft limits was provided by the analysis \cite{Cheung:2021yog} where a universal soft theorem was established in arbitrary theories of scalars (not only Goldstone bosons). The soft limit was formulated in terms of the geometry of the field space (target space). The derivation presented in \cite{Cheung:2021yog,Derda:2024jvo} is very general and also valid in the presence of higher-derivative interactions (in addition to conventional two derivatives) and potential terms. The right-hand side of such generic soft theorems contains derivatives of lower-point amplitudes.

In this paper, we focus on a particular case of two-field sigma models. We show that the phenomenon observed in \cite{Kampf:2019mcd} is absent (it works only for three and more fields), and any theory of Goldstone bosons has an Adler zero. In fact, this property is so stringent that can be used to classify all such models (there are only two of them in the end). Once we enter the non-Goldstone boson territory, naively we get a generic soft theorem of \cite{Cheung:2021yog}. However, we show in this paper that there are special classes of non-Goldstone boson models where the soft theorems are more special and the right-hand side does not contain any derivatives, i.e. can be expressed in terms of lower-point amplitudes (with coefficients depending on the particle multiplicity). The hierarchy of these models is given by the number of (conformal) Killing vectors, and we provide three such examples in this paper, which are all very relevant in the study of sigma models with 2D target space. In fact, our work was motivated by one such model and we indeed found an interesting connection between physical properties and the soft theorems. Our work should open further new directions in the study of sigma models, non-trivial soft theorems and the connection to other modern amplitudes methods.

The paper is organized as follows: In the following subsections, we summarize all the main results. The rest of the paper provides a technical explanation and proof of our statements. In section~\ref{sec2}, we study general sigma models with two Goldston bosons and show that they necessarily have to satisfy Adler zero. In section~\ref{non GB soft}, we discuss the soft theorems for non-Goldstone bosons and introduce our main model. In section~\ref{sec4}, we discuss generalizations including a generic case of non-Goldstone boson theories with no symmetries. Here we derive a soft theorem presenting a special case of the geometric construction \cite{Cheung:2021yog}. In Subsect. \ref{subsec44} we focus on the double soft limit in the models without symmetries. We end the paper with an outlook and future directions.

\subsection{Motivation: sigma models with 2D target space}
\label{ss11}

Our original motivation was to explore classes of non-linear sigma models (NLSM) which recently attracted attention both from the theoretical side \cite{Gamayun:2023sif, Gamayun:2023atu} and in practical applications in condensed matter experiments similar to those discussed in \cite{Batista:2018ctk, Sheu:2024sxf}. The former class is referred to as the Lie-algebraic sigma models \cite{Gamayun:2023sif}. The underlying construction is based in the so-called first-order formalism \cite{Losev:2005pu}. The second class, U(1) fibrated sigma models, has been already studied in the context of the soft-limit theorems in  \cite{Kampf:2019mcd}. It was shown that, instead of the Adler zeros, we arrive at recursion relations in the $SU(N)/SU(N -1)$ sigma models which connect $n$-legs soft-limit amplitudes to a special combination of $(n-1)$-legs amplitudes. These soft recursion relations can be used for the reconstruction of all tree-level amplitudes.  

Of special interest are the Lie-algebraic models with the K\"ahlerian target space generalizing the  $\mathbb{C} P^1$ model. The target space is two-dimensional (2D), i.e. realized on complex fields $\phi$ and $\bar\phi$. The non-linear realization of $sl(2)$ generators has the form 
\begin{eqnarray}
\ell _{1} &=&-\phi ^{2}\,\frac{\partial }{\partial \phi }\,,\quad \ell
_{0}=-\phi \,\frac{\partial }{\partial \phi }\,,\quad \ell _{-1}=-\frac{%
\partial }{\partial \phi }\,  \notag \label{l_i(phi)}
\end{eqnarray}
plus a similar equation for the antiholomorphic algebra. Lie-algebraic $\sigma$ models are obtained as linear combinations of the operators (see below) 
\begin{equation}
\ell _{-1}\overline{\ell }_{-1},\;\ell _{0}\overline{\ell }_{0},\;\ell _{1}%
\overline{\ell }_{1},\;(\ell _{0}\overline{\ell }_{-1}+\mathrm{H.c.}%
),\;(\ell _{0}\overline{\ell }_{1}+\mathrm{H.c.}),\;(\ell _{1}\overline{\ell 
}_{-1}+\mathrm{H.c.})  \label{three}
\end{equation}
with arbitrary coefficients. More precisely, the inverse metric of the target space is expressed as 
\begin{equation}
G^{\phi \overline{\phi }}\,\frac{\partial }{\partial \phi }\frac{\partial }{%
\partial \bar{\phi}}= \sum_{a\bar{b}=1,-1,0}\left( {\mathcal{P}%
}_{a\bar{b}}\,\ell _{a}\bar{\ell}_{\bar{b}}+\mathrm{H.c}\right)  \label{four}
\end{equation}
with a set of numeric coefficients $\{{\mathcal{P}}_{a\bar{b}}\}$. Linear combinations (\ref{three}) span the set of Lie-algebraic metrics $G^{\phi\bar{\phi}}$. So far it was known that the Lie-algebraic structure plays an important role
in loop calculations in two dimensions. As for the soft-limit problem in 4D
tree amplitudes, in this work we managed to identify a two-parametric class of models, which encompasses (and interpolates between) two interesting  representatives of the Lie-algebraic sigma models,
and which presents a non-trivial example, namely
\begin{equation}
{\mathcal{L}}^K_\phi=\partial \bar{\phi}\partial \phi \left[ 1+\alpha (\phi +\bar{%
\phi})\right] ^{K}\,.  \label{five}
\end{equation}
Here $\alpha $ is a real constant which sets the strength of the interaction. The dependence of the amplitudes on $\alpha$ is trivial (power-like), and thus the true parameter which distinguishes different models in this class is $K$. For $K=-2$ we arrive at a 
non-compact versions of $\mathbb{C} P^{1}$ (see below) -- the most
well-known representative of the Lie-algebraic models. Its target
space symmetry is the highest of all Lie-algebraic models, and it exhibits
the Adler zeroes. 
The $K=-1$ case is another example of Lie-algebraic sigma model, which does not obey the Adler zero, nevertheless it exhibits an interesting soft theorem of the relatively simple form, which is different from those allowed for the Goldstone bosons.
The same is true also for general $K$.
The target space in the above model is K\"{a}hler and non-compact.\footnote{%
Similar target spaces emerge as the vacuum manifolds in some ${\mathcal{N}}%
=1 $ 4D super-Yang-Mills theories with a judiciously chosen matter sector.}

Our interest in these models is mainly motivated by the scattering amplitudes perspective and the presence of interesting soft theorems which encode underlying symmetries of the physical models. In addition to deriving and analyzing soft theorems we consider general aspects of two-dimensional K\"ahlerian sigma models (one complex dimension)
not necessarily related to Lie algebras. A typical feature of the models characterized by the Lagrangians ${\mathcal L}=G_{\phi\bar\phi}\, \partial\bar\phi\partial\phi $ is that, as a rule, the target spaces to be considered are not coset spaces
and have reduced symmetries, if at all. In particular, for $K\ne -2$ the Lagrangian (\ref{five}) has only one shift symmetry.
Other generalized $\mathbb{C}P^1$ models from the Lie-algebraic class can have a $U(1)$ isometry \cite{Sheu:2024sxf}.
Under these circumstances a general soft reduction of on-shell amplitudes can be obtained by virtue of 
the geometric method of \cite{Cheung:2021yog}. The correspondent soft-limit theorem requires geometric data of the target space. As is seen from Eq. (3.2) in 
\cite{Cheung:2021yog} we would {\em apriori} need covariant derivatives with respect to the vacuum expectation values of the fields $\phi_{\rm vac}=v$ and $\bar\phi_{\rm vac}=\bar v$
namely, $\nabla_v =\partial_v+\Gamma$ and its complex conjugated (here $\Gamma$ is the Christoffel symbol on the target space). In other words, we would have to know the amplitude as a function of $\phi$ and $\bar \phi$, at least in the vicinity of the point $\{v,\bar v\}$ where we perform the weak field expansion of the Lagrangian. On the other hand, the recursion soft theorem for the model (\ref{five}) with $K=-1$ (for the generic value of $K$ see (\ref{soft_theorem1})) does not need this extra information; our results depend just on the amplitudes themselves through recursion expressions. The same is true also for the ``soft sum rules'' which appear as a consequence of an additional symmetry of the Lagrangian (see section \ref{sec4}).

\subsection{Amplitudes of Goldstone bosons}

Spontaneous symmetry breaking and the Nambu-Goldstone bosons (NGB) play an important role in many physical models, including low-energy sector of QCD, Higgs sector or string theory. NGB scattering amplitudes are long known to vanish in the soft limit, due to the presence of the Adler zero,
\begin{equation}
A_n(k,p_1,p_2,\dots,p_n) \xrightarrow{k\rightarrow 0}  0
\end{equation}
However, this makes certain assumptions about the underlying theory. First, let us review the general form of the tree-level soft theorems for $n$ Goldstone bosons corresponding to some broken symmetry. At the leading order, the dynamics is described by a two-derivative Lagrangian corresponding to a nonlinear sigma model with $n-$dimensional target space and the broken symmetry is realized non-linearly in terms of the Goldstone bosons fields $\phi_{I}$, $I=1,\ldots ,n$. The Goldstone bosons couple to the conserved Noether currents $N^{\mu J}$,
\begin{equation}
\langle 0|N^{\mu J}\left( 0\right) |\phi _{I}\left( p\right) \rangle =-
\mathrm{i}p^{\mu }F_{I}^{J},
\end{equation}
where $F_{I}^{J}$ is a non-degenerate matrix of decay constants.
The soft theorem for Goldstone bosons then reads \cite{Kampf:2019mcd}
\begin{equation}
\lim_{k\rightarrow 0}F_{I}^{J}A_{I\,I_{1}I_{2}\ldots I_{n}}\left(
k,p_{1},p_{2},\ldots p_{n}\right) =-\sum\limits_{j=1}^{n}\mathcal{C}
_{I_{j}K}^{J}A_{I_{1}\ldots I_{j-1}KI_{j+1}\ldots I_{n}}\left(
p_{1},p_{2},\ldots p_{n}\right)  \label{basics1}
\end{equation}
where $A_{I_{1}I_{2}\ldots I_{n}}\left( p_{1},p_{2},\ldots p_{n}\right) $ is the tree-level scattering amplitude of Goldstone bosons and 
\begin{equation}
\mathcal{C}_{IK}^{J} =-\mathcal{C}_{KI}^{J}
\end{equation}
is a fixed set of constants (universal for all $A_{I_{1}I_{2}\ldots I_{n}}$) depending on the theory.


In \cite{Cheung:2014dqa} we constructed an explicit model of Goldstone bosons that follows this soft theorem and also discussed the soft recursion reconstruction of its scattering amplitudes. However, the existence of such Goldstone boson models is only possible if at least 3 fields are present in the theory. Namely, in the case of just two Goldstone bosons, the form of the soft theorem (\ref{basics1}) is so {\it strong}, that it can be used for classification of the possible sigma models. This is done in section \ref{sec2}. The result is that the class of nonlinear sigma models with 2D target space with just two Goldston bosons is simple: there are only two nontrivial models. Both of them are Lie-algebraic and satisfy the Adler zero, (i.e. there is no space for non-trivial soft theorem, as stated above) and correspond to the two versions of the $\mathbb{C}P^1$ model. These are described by the Lagrangian (up to admissible field redefinition)
\begin{equation}
\mathcal{L}_{\pm }=\frac{\partial \overline{\phi}\cdot
\partial \phi }{\left( 1\pm \alpha ^{2}\overline{\phi}\phi \right) ^{2}}
\label{CP1 Lagrangian 1}.
\end{equation}
and are related with the target space $S^{2}\approx O(3)/O(2)$ and its non-compact version with the target space $H^{2}\approx O(1,2)/O(2)$.
Stated differently, the Lagrangian of any non-linear sigma model describing just two Goldstone bosons, i.e. with two-dimensional target space which is a coset ${\cal{G}}/{\cal{H}}$, can be transformed to (\ref{CP1 Lagrangian 1}) by admissible field redefinition\footnote{Of course, such a classification could be obtained independently by means of analysis of the underlying geometry. In our proof we use the amplitude methods in order to stress the relation of the problem to the soft theorems.}.   

%
%

\subsection{New non-Goldstone model and non-trivial soft theorems\label{nngm}}

There is a natural hierarchy of non-linear sigma models based on the right-hand side of the associated soft theorems. Based on the results of \cite{Kampf:2019mcd} and \cite{Cheung:2021yog} we get
$$
\includegraphics[scale=0.37]{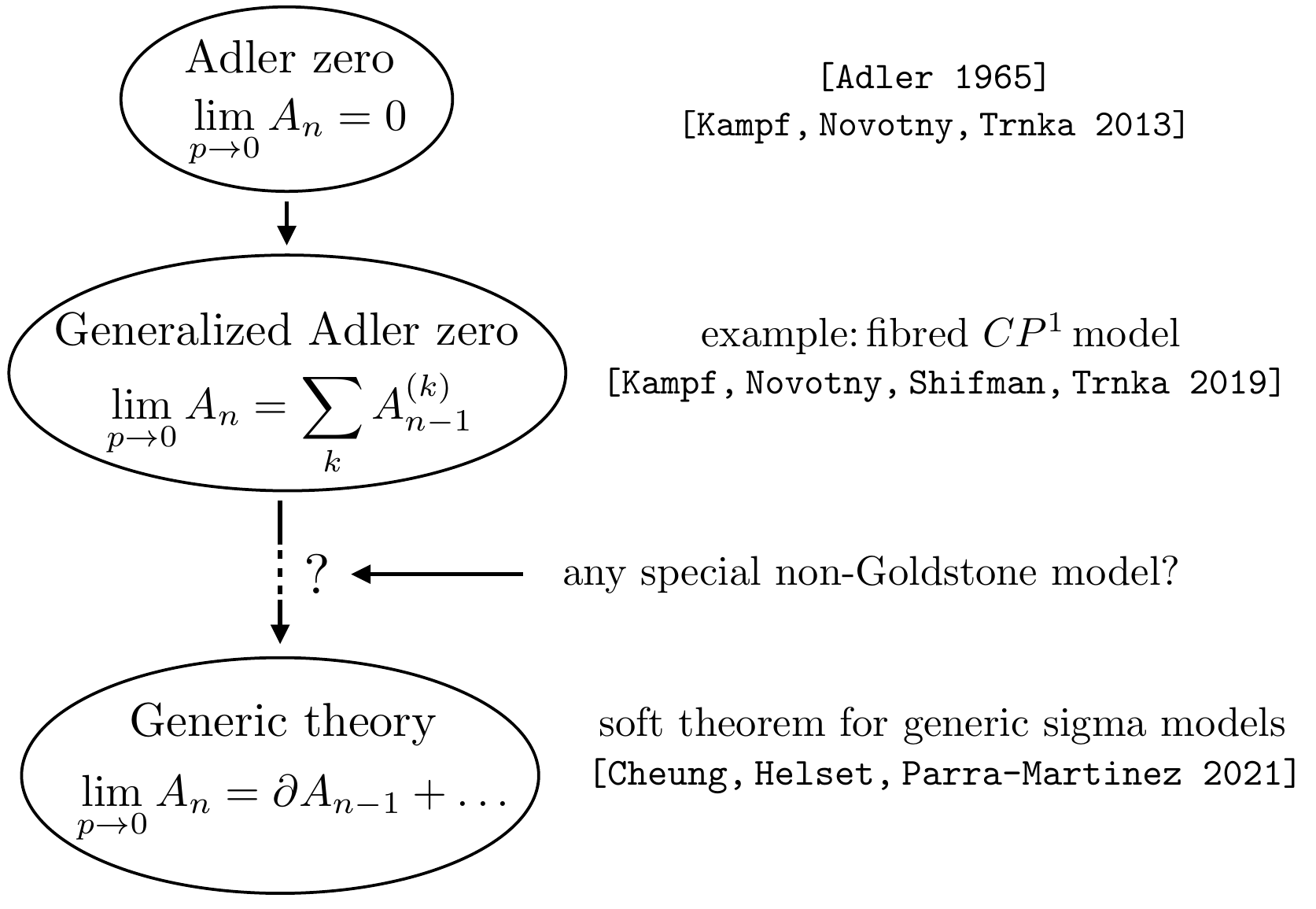}
$$
where the first two levels correspond to the Goldstone bosons (with or without the presence of Adler zero), while the bottom level is a generic non-linear sigma model case with a soft theorem found in \cite{Cheung:2021yog}. It is an open question if there are any special non-NGB models with better-than-generic soft theorems, namely with the absence of the derivatives of the amplitude. In this paper, we give an answer to the two-scalar case. As discussed above all theories with Goldstone bosons do have Adler zero. 
The hierarchy of theories is simplified,
$$
\includegraphics[scale=0.4]{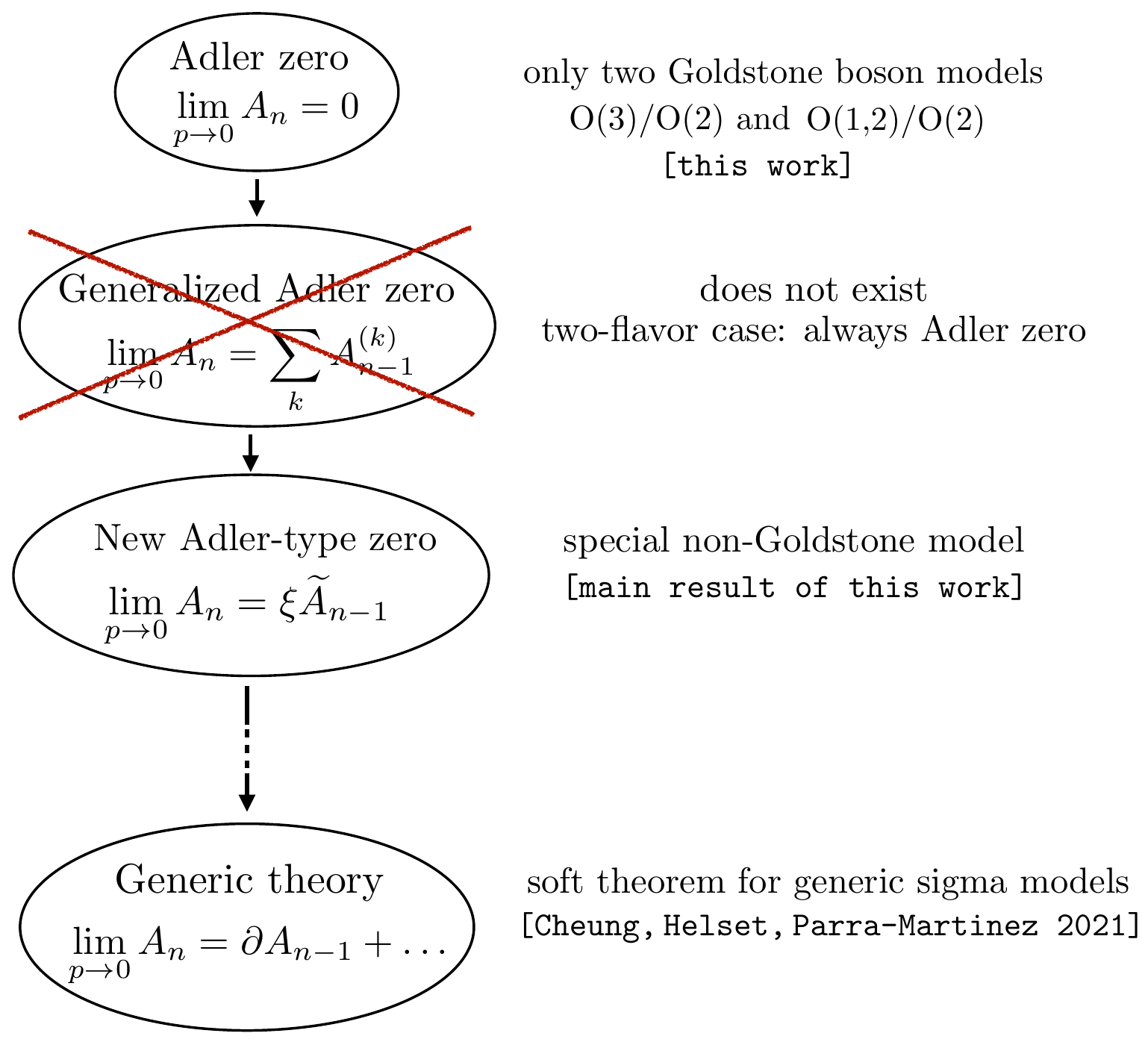}
$$
and the aim of this paper is to show there are non-Goldstone theories with special soft theorems. Inspired by the study of sigma models with 2D target space we come up with a two-parametric  model given by the Lagrangian (\ref{five}),
\begin{equation}
\mathcal{L}_{\phi }^{K}=\partial \overline{\phi }\cdot \partial \phi \left(
1+\alpha \left( \phi +\overline{\phi }\right) \right) ^{K} \,. \label{aK}
\end{equation}
As stated above and discussed further in more detail in section \ref{subsec31}, this model contains the non-compact variant of the $\mathbb{C}P^1$ model as the special case for $K=-2$ and another Lie-algebraic sigma model for $K=-1$. The Lagrangian (\ref{aK}) is a generalization of both theories and interpolates between them. We denote two types of fields $\pm$ in scattering amplitudes. We will prove in section \ref{subsec32} that the tree-level soft theorem takes a form,
\begin{equation}
\lim_{p\rightarrow 0}A_{n^{+}+1,n^{-}}\hspace{-0.1cm}\left(p^{+},1^{+},{\dots},n^{+},1^{-},{\dots} ,n^{-}\right) = \xi_{\alpha,K}^{(n^+,n^-)} \cdot A_{n^{+},n^{-}}\hspace{-0.1cm}\left( 1^{+},{\dots}
,n^{+},1^{-},{\dots},n^{-}\right)  \label{soft_theorem1}
\end{equation}
where the coefficient $\xi$ is given by 
\begin{equation}
\xi_{\alpha,K}^{(n^+,n^-)} = \frac{1}{2}\alpha \left[\left( K{+}2\right)
(2{-}n^{+}{-}n^{-}) {-} K\left( n^{+}{-}n^{-}\right)\right]
\end{equation}
and similarly for $p^{-}\rightarrow 0$, with the exchange $n^{+}\leftrightarrow n^{-}$. For the maximal symmetry limit $K\rightarrow -2$ we recover the Adler zero (note that in this case, only the amplitudes with 
$n^{+}=n^{-}$ are nonzero). For the second Lie-algebraic sigma model for $K=-1$ we get 
\begin{eqnarray}
&&\lim_{p\rightarrow 0}A_{n^{+}+1,n^{-}}\hspace{-0.1cm}\left(p^{+},1^{+},{\dots},n^{+},1^{-},{\dots} ,n^{-}\right) 
\nonumber\\[1mm]
&&= \alpha(1{-}n_{-}) \cdot A_{n^{+},n^{-}}\hspace{-0.1cm}\left( 1^{+},{\dots}
,n^{+},1^{-},{\dots},n^{-}\right) . 
\label{soft_theorem2}
\end{eqnarray}
Note that these soft theorems are different from the ones found for the Goldstone boson amplitudes in \cite{Kampf:2019mcd} -- the right-hand side is again simply a lower point amplitude with a soft leg removed, but there is a non-universal prefactor (depending on the particle multiplicities) in front of the amplitude. We provide an explicit example and verification of the validity of the soft theorem for 5pt amplitudes in Appendix \ref{appB}.

Similarly, the general geometric double soft theorems discussed in \cite{Cheung:2021yog} contain the derivatives of the amplitudes with respect to the vacuum expectation value.
Our model based on the Lagrangian (\ref{five}) does not need such a type of information. We discuss this aspect in section \ref{subsec44}, where the explicit form of the various variants of the double soft theorems can be found.

\subsection{More general models and soft theorem sum rules}

The model discussed above was significant from the symmetry perspective. The Lagrangian (in appropriately chosen coordinates) was invariant with respect to the infinitesimal shift transformations, and transformed in a special way with respect to scaling and special conformal transformations of the target space.  This gave rise to an algebra of (conformal) Killing vectors corresponding to $SU(1,1)\approx SO(1,2)$. We discuss the details in section \ref{subsec31}. The generalization to a larger class of models with the general Lagrangian 
\begin{equation}
\mathcal{L}=G\left( \phi ,\overline{\phi }\right) \partial \phi \cdot
\partial \overline{\phi }  
\end{equation}
where $G(\phi,\overline{\phi})$ is the target space metric, uses the same ingredient: the existence of appropriate (conformal) Killing vector on the target space, associated with a current which couples to some superposition of one-particle states. We provide two examples of a wider class of such non-linear sigma models. 

\subsubsection*{Homogeneous models}

The first example deals with the target space metric with just one special conformal
Killing vector. 
Suppose, that the function
 $G\left( z,\overline{z}\right) $ is homogeneous of degree $K$, i.e. $ G\left( \lambda \phi ,\lambda \overline{\phi }\right) =\lambda ^{K}G\left(\phi ,\overline{\phi }\right)$. 
 The desired conformal Killing vector 
 $$\ell _{0}=\phi \partial _{\phi }+\overline{\phi }\partial _{\overline{\phi }}$$
 corresponds to the infinitesimal scaling $\delta \phi =\phi$, $\delta\overline{\phi }=\overline{\phi }$. Performing the weak field expansion $\phi(x)=w+z(x)$ around the vacuum point $\langle \phi\rangle=w$ we get an expansion for the metric,
\begin{equation}
G(\phi,\overline{\phi}) = G\left( z+w,\overline{z}+\overline{w}\right) =  g+z\partial _{w}g+\overline{z}
\partial _{\overline{w}}g+\ldots \quad \mbox{where \,\,$g=G\left( w,\overline{w}\right)$}
\end{equation}
While a scattering amplitude in this model does not satisfy a simplified soft theorem, we can nevertheless derive ``soft sum rule'' in the form of the soft limit of a linear combination of amplitudes with different ``flavors'' of the soft
particle, 
\begin{align}
&\lim_{p\rightarrow 0}g^{1/2}\left( \overline{w} A_{n^{+}+1,n^{-}} + wA_{n^{+},n^{-}+1}\right) \\
&\hspace{1.5cm} =\frac{1}{2}\left[ \left( n^{+}{-}n^{-}\right) \left( g^{-1}w\partial
_{w}g{-}g^{-1}\overline{w}\partial _{\overline{w}}g\right) {+}(K{+}2)\left(
2{-}n^{+}{-}n^{-}\right) \right] \cdot A_{n^{+},n^{-}} \nonumber
\end{align}
 Note that this sum rule does not contain derivatives of the amplitudes on the right-hand side, but again just a prefactor multiplied by a lower-point amplitude. In this case the prefactor does depend on the metric (and its derivatives) of the target space and the multiplicity of the particles. For the proof and further discussion see section \ref{subsec41}.

\subsubsection*{$U(1)$ invariant models}

In our second example, the target space metric has a special form and only depends on the product $\phi\overline{\phi}$, i.e. $G(\phi\overline{\phi})$ where $G$ is real. The Lagrangian is then invariant with respect to the $U\left( 1\right) $ rotations 
\begin{equation*}
\phi \rightarrow \mathrm{e}^{\mathrm{i}\alpha }\phi ,~~~\overline{\phi }%
\rightarrow \mathrm{e}^{-\mathrm{i}\alpha }\overline{\phi }\,.
\end{equation*}
and there is an associated conserved current which couples to the combination of states $w|p^{+}\rangle -\overline{w}|p^{-}\rangle$ where $w,\overline{w}$ are again vacuum parameters in the weak field expansion. As a result, we can derive a ``soft sum rule'' in the form
\begin{equation}
\lim_{p\rightarrow 0}g^{1/2}\left( wA_{n^{+},n^{-}+1}-\overline{w}
A_{n^{+}+1,n^{-}}\right) =\left( n^{+}-n^{-}\right) \left( 1+w\overline{w}
g^{-1}g^{\prime }\right) A_{n^{+},n^{-}}\,.  \label{soft sum rule U(1) 1}
\end{equation}
This is a generic case for arbitrary $w,\overline{w}$. There are special cases such as $w=\overline{w}=0$, where only amplitudes with $n^+=n^-$ do not vanish. Then Eq. (\ref{soft sum rule U(1) 1}) degenerates into a trivial $0=0$ equality.
This is the case of the  ${\mathbb C}P^1$ models with Lagrangian (\ref{CP1 Lagrangian 1}).

\section{Sigma models for two Goldstone bosons in detail}
\label{sec2}

We start our discussion with a special case when the sigma model describes the leading order low-energy dynamics of just two Goldstone bosons corresponding to some broken symmetry. For the Goldstone bosons, the soft behavior of the scattering amplitudes is well understood \cite{Cheung:2014dqa}. We will show that in the case of just two Goldstone bosons, the constraints due to the soft theorem are so strong, that we are able to completely classify the possible theories, their perturbative $S$-matrices and target spaces of the corresponding non-linear sigma models.

\subsection{Soft theorems for Goldstone bosons}
\label{subsec21}

Let us first remind the general form of the tree-level soft theorems for $n$ Goldstone bosons corresponding to some broken symmetry.\footnote{%
For more details and explicit examples see \cite{Cheung:2014dqa}.} At the leading
order, the dynamics is described by means of two-derivative Lagrangian
corresponding to a nonlinear sigma model with $n-$dimensional target space
and the broken symmetry is realized non-linearly in terms of the Goldstone
bosons fields $\phi _{I}$, $I=1,\ldots ,n$. The Goldstone bosons couple to
the conserved Noether currents $J^{\mu J}$,
\begin{equation}
\langle 0|J^{\mu\, J}\left( 0\right) |\phi _{I}\left( p\right) \rangle =-
\mathrm{i}p^{\mu }F_{I}^{J},
\end{equation}
where $F_{I}^{J}$ is a nondegenerate matrix of decay constatnts. Assume,
that these currents are expressed in terms of the Goldstone boson fields $\phi _{I}$ as\,\footnote{%
Here and in what follows, we tacitly assume summation over repeating indices.%
}
\begin{equation}
J^{J}=F_{I}^{J}\partial \phi _{I}+\mathcal{K}_{IK}^{J}\,\phi _{K}\partial \phi
_{I}+O\left( \phi ^{3}\right) ,
\label{fk22}
\end{equation}
where $\mathcal{K}_{IK}^{J}$ is a set of constant coefficients. The soft
theorem for Goldstone bosons reads then \cite{Cheung:2014dqa}
\begin{equation}
\lim_{k\rightarrow 0}F_{I}^{J}A_{I~I_{1}I_{2}\ldots I_{n}}\left(
k,p_{1},p_{2},\ldots p_{n}\right) =-\sum\limits_{j=1}^{n}\mathcal{C}
_{I_{j}K}^{J}A_{I_{1}\ldots I_{j-1}KI_{j+1}\ldots I_{n}}\left(
p_{1},p_{2},\ldots p_{n}\right)  \label{basics2}
\end{equation}
where $A_{I_{1}I_{2}\ldots I_{n}}\left( p_{1},p_{2},\ldots p_{n}\right) $ is
the tree-level scattering amplitude of the Goldstone bosons $\phi
_{I_{i}}(p_{i}) $ (we treat them all as being out-going), and the coefficients
$\mathcal{C}_{IK}^{J}$ are as follows,
\begin{equation}
\mathcal{C}_{IK}^{J}=\frac{1}{2}\left( \mathcal{K}_{IK}^{J}-\mathcal{K}
_{KI}^{J}\right) =-\mathcal{C}_{KI}^{J},
\end{equation}
with $\mathcal{K}_{IK}^{J}$ defined in Eq. (\ref{fk22}).
There are two sources of non-vanishing $\mathcal{C}_{IK}^{J}$ which can be
traced back to the Lagrangian formulation: either there is a cubic vertex
and/or the transformation of the fields $\phi _{I}$ with respect to the
broken symmetry contains a term linear in $\phi _{I}$. The formula (\ref
{basics2}) generalizes the Adler zero.

For a violation of the Adler zero, i.e. for validity of nontrivial soft
theorem with nonzero right-hand side, we need the following two conditions
to be satisfied

\begin{itemize}
\item both even-particle and left-hand amplitudes are nonzero

\item treating the amplitudes as tensors with indices $I_{1}I_{2}\ldots
I_{n} $, this tensor must not be invariant with respect to the linear
transformation induced by the matrix $\mathcal{C}_{IK}^{J}$.
\end{itemize}

In the case of just two Goldstone bosons (i.e. when the target space of the
nonlinear sigma model is two-dimensional), the form of $\mathcal{C}
_{IK}^{J} $ is simply 
\begin{equation*}
\mathcal{C}_{IK}^{J}=c^{J}\varepsilon _{IK}
\end{equation*}
where $\varepsilon _{IK}$ is the two-dimensional Levi-Civita symbol and 
\begin{equation*}
c^{J}=\frac{1}{2}\varepsilon _{IK}\mathcal{C}_{IK}^{J}.
\end{equation*}
Note that $\varepsilon _{IK}$ is a generator of $SO(2)\approx U\left(
1\right) $. Therefore, to get a nontrivial soft theorem with a nonzero right-hand side, the $S$-matrix cannot be invariant with respect to $SO(2)$, i.e.
the $U\left( 1\right) $ charge cannot be conserved. Of course, the
manifestly $U\left( 1\right) $ invariant Lagrangians automatically imply
Adler zero since they produce only even particle amplitudes. However, in the
general case the $U\left( 1\right) $ symmetry could be hidden and realized
non-linearly at the Lagrangian level, although leading to the $U\left(
1\right) $ invariant $S$-matrix. In fact, as we will prove in the next
subsection, it is always the case -- there is no nontrivial soft theorem in
the case of non-linear sigma models with two Goldstones. As a consequence,
this means that for any non-linear sigma model with Lagrangian describing two
Goldstone bosons, the $S$-matrix possesses a hidden $U\left( 1\right) $ symmetry.

\subsection{No-go theorem for Adler zero violation in two-Goldstone boson
non-linear sigma models}\label{subsec22}

In this subsection, we will use amplitude methods which are very efficient for our purpose, since they are indpendent on the form of the Lagrangian and the coordinates choice on the target space. Suppose that we have some general sigma model with a two-dimensional target space describing two Goldstone bosons (let us denote them $\psi $ and $\chi $). Then the tree-level amplitudes satisfy the nontrivial soft theorem of the general form (\ref{basics2}), which can be written as\,%
\footnote{Here we use a shorthand notation for the momenta $p_{i}\equiv i$. The
subscripts $\psi ,\chi $\ denote the type of the Goldstone boson. 
Without loss of generality we put $F_{I}^{J}\to \delta _{I}^{J}$ using the fact that we can make an appropriate linear
transformation of the currents and the one-particle states.}
\begin{eqnarray}
&&\lim_{n+1\rightarrow 0}A_{n+1-m,m}\left( 1_{\chi },\ldots ,m_{\chi
},\left( m+1\right) _{\psi },\ldots ,\left( n+1\right) _{\psi }\right) 
\notag \\
&=&-C\sum\limits_{j=1}^{m}A_{n-m+1,m-1}\left( 1_{\chi },\ldots ,j_{\psi
},\ldots m_{\chi }\left( m+1\right) _{\psi },\ldots ,n_{\psi }\right)  \notag
\\
&&+C\sum\limits_{i=m+1}^{n}A_{n-m-1,m+1}\left( 1_{\chi },\ldots ,m_{\chi
},\left( m+1\right) _{\psi },\ldots ,i_{\chi },\ldots ,n_{\psi }\right) ,
\label{soft theorem a}
\end{eqnarray}
and
\begin{eqnarray}
&&\lim_{1\rightarrow 0}A_{n-m,m}\left( 1_{\chi },\ldots ,m_{\chi },\left(
m+1\right) _{\psi },\ldots ,n_{\psi }\right)  \notag \\
&=&-D\sum\limits_{j=2}^{m}A_{n+1-m,m-2}\left( 2_{\chi },\ldots ,j_{\psi
},\ldots m_{\chi }\left( m+1\right) _{\psi },\ldots ,n_{\psi }\right)  \notag
\\
&&+D\sum\limits_{i=m+1}^{n}A_{n-m,m-1}\left( 2_{\chi },\ldots ,m_{\chi
},\left( m+1\right) _{\psi },\ldots ,i_{\chi },\ldots ,n_{\psi }\right) ,
\label{soft theorem b}
\end{eqnarray}
with $C,D$ fixed. Let us now show that the only possibility is that all
the left-hand amplitudes vanish.

The 3pt amplitudes vanish on shell, therefore let us start with the 4pt and 5pt amplitudes. These are contact, and the power counting and permutational symmetry allow only for the following forms,
\begin{eqnarray}
A_{4,0}\left( 1_{\psi },2_{\psi },3_{\psi },4_{\psi }\right)
&=&A_{0,4}\left( 1_{\chi },2_{\chi },3_{\chi },4_{\chi }\right)
=A_{3,1}\left( 1_{\psi },2_{\psi },3_{\psi },4_{\chi }\right) = A_{1,3}\left( 1_{\psi },2_{\chi },3_{\chi },4_{\chi }\right) =0\,,  \notag\\
A_{2,2}\left( 1_{\psi },2_{\psi },3_{\chi },4_{\chi }\right)
&=&a_{2,2}\left( 1+2\right) ^{2}=2a_{2,2}\left( 1\cdot 2\right)
=2a_{2,2}\left( 3\cdot 4\right).  \label{4pt ansatz}
\end{eqnarray}
Here we use a following notation for the momenta:
\begin{equation}
(a\cdot b) = p_a\cdot p_b, \quad (a+b+c)^2 = (p_a+p_b+p_c)^2\quad\mbox{etc.}
\end{equation}
The non-vanishing 4pt amplitude satisfies the soft theorem trivially since it has manifest Adler zero in all channels and since the 3pt amplitudes vanish on shell. Similarly, the most general ansatz for the 5pt amplitudes reads
\begin{eqnarray}
A_{5,0}\left( 1_{\psi },2_{\psi },3_{\psi },4_{\psi },5_{\psi }\right)
&=&A_{0,5}\left( 1_{\chi },2_{\chi },3_{\chi },4_{\chi },5_{\chi }\right)
=A_{4,1}\left( 1_{\psi },2_{\psi },3_{\psi },4_{\psi },5_{\chi }\right) 
\notag\\
&=&A_{1,4}\left( 1_{\psi },2_{\chi },3_{\chi },4_{\chi },5_{\chi }\right) =0\,,
\notag \\
A_{2,3}\left( 1_{\psi },2_{\psi },3_{\chi },4_{\chi },5_{\chi }\right)
&=&a_{2,3}(1+2)^{2}=a_{2,3}\left( 3+4+5\right)^{2}\,,  \notag \\
A_{3,2}\left( 1_{\psi },2_{\psi },3_{\psi },4_{\chi },5_{\chi }\right)
&=&a_{3,2}(1+2+3)^{2}=a_{3,2}\left( 4+5\right)^{2} \,. \label{5pt ansatz}
\end{eqnarray}
Here the soft theorem is trivial for $A_{5,0}$ and $A_{0,5}$ since $%
A_{3,1}=A_{1,3}=0$, and for $A_{4,1}$ and $A_{1,4}$ since e.g. for $A_{4,1}$
the soft theorem (\ref{soft theorem a}) for $1\rightarrow 0$ is 
\begin{eqnarray}
&&\lim_{1\rightarrow 0}A_{4,1}\left( 1_{\psi },2_{\psi },3_{\psi },4_{\psi
},5_{\chi }\right) = CA_{2,2}\left( 2_{\chi },3_{\psi },4_{\psi },5_{\chi }\right)
+CA_{2,2}\left( 2_{\psi },3_{\chi },4_{\psi },5_{\chi }\right)  \notag\\
&&\hspace{4cm} +CA_{2,2}\left( 2_{\psi },3_{\psi },4_{\chi },5_{\chi }\right)
-CA_{4,0}\left( 2_{\psi },3_{\psi },4_{\psi },5_{\psi }\right)  \notag\\
&&\hspace{4cm} =2Ca_{2,2}\left[ \left( 2\cdot 5\right) +\left( 2\cdot 4\right) +\left(
2\cdot 3\right) -0\right] =0
\end{eqnarray}
while for $5\rightarrow 0$ we get (\ref{soft theorem b}) in the form
\begin{eqnarray}
&&\lim_{5\rightarrow 0}A_{4,1}\left( 1_{\psi },2_{\psi },3_{\psi },4_{\psi
},5_{\chi }\right)=DA_{3,1}\left( 1_{\chi },2_{\psi },3_{\psi },4_{\psi }\right)
+DA_{3,1}\left( 1_{\psi },2_{\chi },3_{\psi },4_{\psi }\right)  \notag \\
&&\hspace{3cm} +DA_{3,1}\left( 1_{\psi },2_{\psi },3_{\chi },4_{\psi }\right)
+DA_{3,1}\left( 1_{\psi },2_{\psi },3_{\psi },4_{\chi }\right) =0\,,
\end{eqnarray}
and similarly for $A_{4,1}$. For $A_{2,3}$ we get directly
\begin{equation}
\lim_{1\rightarrow 0}A_{2,3}\left( 1_{\psi },2_{\psi },3_{\chi },4_{\chi
},5_{\chi }\right) =\lim_{1\rightarrow 0}a_{2,3}(1+2)^{2}=0
\end{equation}
while the right-hand side of the soft theorem (\ref{soft theorem a}) gives
\begin{eqnarray}
&&\lim_{1\rightarrow 0}A_{2,3}\left( 1_{\psi },2_{\psi },3_{\chi },4_{\chi
},5_{\chi }\right) 
=CA_{0,4}\left( 2_{\chi },3_{\chi },4_{\chi },5_{\chi }\right)
-CA_{2,2}\left( 2_{\psi },3_{\psi },4_{\chi },5_{\chi }\right)  \notag \\
&&\hspace{5cm} -CA_{2,2}\left( 2_{\psi },3_{\chi },4_{\psi },5_{\chi }\right)
+CA_{2,2}\left( 2_{\psi },3_{\chi },4_{\chi },5_{\psi }\right)  \notag \\
&&\hspace{3.5cm} =2C\left[ 0-a_{2,2}\left( 2\cdot 3\right) -a_{2,2}\left( 2\cdot 4\right)
-a_{2,2}\left( 2\cdot 5\right) \right] =0,
\end{eqnarray}
and thus the soft theorem holds without any constraint. On the other hand
\begin{equation}
\lim_{5\rightarrow 0}A_{2,3}\left( 1_{\psi },2_{\psi },3_{\chi },4_{\chi
},5_{\chi }\right) =a_{2,3}(1+2)^{2}
\end{equation}
which is nonzero for general kinematics, while the soft theorem (\ref{soft
theorem b}) requires
\begin{eqnarray}
&&\lim_{5\rightarrow 0}A_{2,3}\left( 1_{\psi },2_{\psi },3_{\chi },4_{\chi
},5_{\chi }\right)  = DA_{1,3}\left( 1_{\chi },2_{\psi },3_{\chi },4_{\chi }\right)
+DA_{1,3}\left( 1_{\psi },2_{\chi },3_{\chi },4_{\chi }\right)  \notag \\
&&\hspace{3cm} -DA_{3,1}\left( 1_{\psi },2_{\psi },3_{\psi },4_{\chi }\right)
-DA_{3,1}\left( 1_{\psi },2_{\psi },3_{\chi },4_{\psi }\right) =0\,.
\end{eqnarray}
This is compatible only for $a_{2,3}=0$. Similarly, we get $a_{3,2}=0$ as a consequence of soft theorem (\ref{soft theorem a}). Thus, all the 5pt amplitudes vanish. This implies, using again the soft theorem, that the 6pt amplitudes have the Adler zero in both $\chi $ and $\psi $ channels.

Let us take $n>2$ and suppose now, that all odd amplitudes up to and
including $\left( 2n-1\right) $pt vanish. According to the assumed soft theorem, the $\left( 2n\right) $pt amplitudes have then the Adler zero in both $\chi $ and $\psi $ channels. As above, the $\left( 2n+1\right) $pt amplitudes have to be contact (because there is no nontrivial factorization channel as a consequence of the assumption) and of the form 
\begin{eqnarray}
&&A_{m,2n-m+1}(1_{\psi },\ldots ,m_{\psi },m+1_{\chi },\ldots ,\left(
2n+1\right) _{\chi })  \\
&&\hspace{1cm} =a_{m,2n-m+1}\left( 1+2+\ldots +m\right) ^{2} = a_{m,2n-m+1}\left( \left( m+1\right) +\ldots \left( 2n+1\right) \right)  
^{2}. \notag
\end{eqnarray}
Note that $A_{0,2n+1}=A_{1,2n}=A_{2n+1,0}=A_{2n,1}=0$ from power counting,
permutational symmetry and kinematics. Let us take the amplitude $A_{m,2n-m+1}$ with $m>3$ and $2n-m>0$ and calculate the successive double soft limit
\begin{eqnarray}
&&\lim_{2\rightarrow 0}\left[ \lim_{1\rightarrow 0}A_{m,2n-m+1}(1_{\psi
},\ldots ,m_{\psi },\left( m+1\right) _{\chi },\ldots ,\left( 2n+1\right)
_{\chi })\right]  \\[1mm]
&&\hspace{1cm}= a_{m,2n-m+1}\left( 3+\ldots +m\right) ^{2}  = a_{m,2n-m+1}\left( \left( m+1\right) +\ldots +\left( 2n+1\right) \right)
^{2}\,,\notag
\end{eqnarray}
which is nonzero for generic momenta. On the other hand using the soft
theorem for the $1\rightarrow 0$ limit and then the Adler zero for $\left(
2n\right) $pt amplitudes for the $2\rightarrow 0$ limit, we get
\begin{equation}
\lim_{2\rightarrow 0}\left[ \lim_{1\rightarrow 0}A_{m,2n-m+1}(1_{\psi
},\ldots ,\left( m\right) _{\psi },\left( m+1\right) _{\chi },\ldots ,\left(
2n+1\right) _{\chi })\right] =0
\end{equation}
and thus $a_{m,2n-m+1}=0$ for $m=4,\ldots ,2n-1$. For the amplitudes $
A_{2,2n-1}$ and $A_{3,2n-2}$ we can similarly calculate the limit
\begin{eqnarray}
\lim_{2n\rightarrow 0}\left[ \lim_{2n+1\rightarrow 0}A_{m,2n-m+1}\right]
&=&a_{m,2n-m+1}\left( 1+\ldots +m\right) ^{2}  \notag \\
&=&a_{m,2n-m+1}\left( \left( m+1\right) +\ldots +\left( 2n-1\right) \right)
^{2}
\end{eqnarray}
where $m=2,3$. Using again the soft theorem for $2n+1\rightarrow 0$ and
Adler zero for $\left( 2n\right) $pt amplitudes for $2n\rightarrow 0$ we
conclude $A_{2,2n-1}=A_{3,2n-2}=0$. By induction, all the $\left(
2n+1\right) $pt amplitudes then vanish.

As a result, provided the theory satisfies the soft theorem (\ref{soft
theorem a}), (\ref{soft theorem b}), there are only even amplitudes in
the theory and they obey the Adler zero conditions in both $\chi $ and $\psi $
channels. Moreover, provided $C,D\neq 0$, the soft theorem applied to the
vanishing odd amplitudes ensures the $SO\left( 2\right) $ invariance of the 
$S$-matrix, i.e. the charge conservation.

\subsection{The S-matrix reconstruction}\label{subsec23}

The soft theorem of the type (\ref{basics2}) is known to determine the
tree-level $S$-matrix uniquely in terms of the set of lowest point seed
amplitudes \cite{Cheung:2014dqa}. For the two-derivative sigma models, the seed
amplitudes are the 4pt ones. The reconstruction of the tree-level  $S-$%
matrix can be performed using the soft Britto-Cachazo-Feng-Witten (BCFW) recursion which is based on appropriate one-parametric shift of the particle momenta which allows to
probe the soft limits of the $n$pt tree amplitude $A_{n}\left( p_{1},\ldots
,p_{n}\right) $, 
\begin{eqnarray}
\widehat{p}_{i} &=&p_{i}\left( 1-a_{i}z\right) ,~~~i=1,\ldots ,n-2\,,  \notag \\[1mm]
\widehat{p}_{j} &=&p_{j}+zq_{j},~~~~j=n-1,n\,,
\end{eqnarray}
where $z$ is the shift parameter, $a_{i}\neq a_{j}$ are set of constants
and $q_{j}$ satisfy the constraints stemming from momentum conservation and
the on-shell conditions for the shifted momenta. The amplitudes then become
functions of $z$, 
\begin{equation}
A_{n}\left( z\right) \equiv A_{n}\left( \widehat{p}_{1},\ldots ,\widehat{p}%
_{n}\right) .
\end{equation}
Then $A_{n}\left( 0\right) =A_{n}\left( p_{1},\ldots ,p_{n}\right) $ and
according to the power counting,\footnote{%
Note that we assume two-derivative Lagrangians.} $A_{n}\left( z\right)
=O\left( z^{2}\right) $ for $z\rightarrow \infty $. \ This asymptotic
behavior ensures, that the meromorphic function
\begin{equation}
f_{n}\left( z\right) =\frac{A_{n}\left( z\right) }{z\prod\limits_{i=1}^{n-2}
\left( 1-a_{i}z\right) }
\end{equation}
has  vanishing residue at infinity for $n>4$. Applying the residue theorem to 
$f_{n}\left( z\right) $ we can express the amplitude as a sum over residues
of $f_{n}\left( z\right) $ at factorization poles $z_{k}$ and soft poles $
1/a_{j}$
\begin{equation}
A_{n}\left( p_{1},\ldots ,p_{n}\right) =-\sum\limits_{k}\mathrm{res}\left(
f_{n},z_{k}\right) -\sum\limits_{l}\mathrm{res}\left( f_{n},\frac{1}{a_{l}}
\right) .
\end{equation}
The both corresponding residues can be expressed in terms of lower point
amplitudes, using either the factorization theorem or the soft theorem. This
allows for recursive reconstruction of the full set of amplitudes, starting
with the 4pt ones as the input.

According to the results of the previous subsection, any non-linear sigma
model describing the low-energy dynamics of just two Goldstone bosons,
generates only even point amplitudes which satisfy the Adler zero in both
channels. Therefore, the soft recursion described above allows for
reconstruction of the complete tree-level $S$-matrix in terms of just one
free parameter which can be identified with the constant $a_{2,2}$ defining
the only non-vanishing 4pt amplitude (cf. Eq. (\ref{4pt ansatz})). This
reconstruction is unique.

On the other hand, at least two examples of two-Goldstone boson nonlinear sigma models are known. These are the $\mathbb{C}P^{1}$ model corresponding 
to the target space $S^{2}\approx O(3)/O(2)$ and
its non-compact version with the target space $H^{2}\approx O(1,3)/O(2)$ and the
Lagrangian
\begin{equation}
\mathcal{L}_{\pm }=\frac{1}{2}\frac{\left( \partial \psi \right)^{2}+\left(
\partial \chi \right)^{2}}{\Big( 1\pm \frac{1}{2}\alpha ^{2}\left( \psi
^{2}+\chi ^{2}\right) \Big) ^{2}}=\frac{\partial \overline{\phi }\cdot
\partial \phi }{\left( 1\pm \alpha ^{2}\overline{\phi }\phi \right) ^{2}}
\label{CP1 Lagrangian}
\end{equation}
where we introduced complex fields
\begin{equation*}
\phi ,\overline{\phi }=\frac{1}{\sqrt{2}}\left( \chi \pm \mathrm{i}\psi
\right) .
\end{equation*}
In the complex coordinates, the $U\left( 1\right) $ symmetry is manifest.
Moreover, these models are special examples of the Lie-algebraic sigma models
with inverse metric
\begin{equation*}
G^{\phi \overline{\phi }}\partial _{\phi }\partial _{\overline{\phi }}=\ell
_{-1}\overline{\ell }_{-1}\pm 2\alpha ^{2}\ell _{0}\overline{\ell }
_{0}+\alpha ^{4}\ell _{1}\overline{\ell }_{1}
\end{equation*}
with $\ell _{i}$ and $\overline{\ell }_{i}$  given by (\ref{l_i(phi)}).
Both these models satisfy the Adler zero, and the seed 4pt amplitude reads
\begin{equation*}
A_{2,2}\left( 1_{\psi }2_{\psi }3_{\chi }4_{\chi }\right) =\pm 4\alpha
^{2}\left( 1+2\right) ^{2}
\end{equation*}
and thus $a_{2,2}=\pm 4\alpha ^{2}\,.$  This covers all possibilities for the general case discussed above. Since the soft BCFW recursion is unique, we
can conclude, that the only nontrivial non-linear sigma models for just two
Goldstones are the above two models with target spaces $O(3)/O(2)$ and $%
O(1,2)/O(2)$ with Lagrangian (\ref{CP1 Lagrangian}) (up to an admissible
field redefinition).

\section{Soft theorems for non-Goldstone bosons\label{non GB soft}}

According to the results of the previous section, provided both elementary excitations described by nonlinear sigma model with 2D target space are Goldstone bosons, the tree-level amplitudes satisfy the Adler zero and the $S-$matrix possess $U\left( 1\right) $ symmetry. Stated differently, provided there exist two nonlinearly realized symmetries of the Lagrangian, the Noether currents of which couple to the one-particle states, the soft theorem is trivial with zero right-hand side and the Lagrangian is automatically invariant with respect to additional $U\left( 1\right)$ symmetry being (up to field redefinition) identical to that of $O(3)/O(2)$
and $O(1,2)/O(2)$ models (\ref{CP1 Lagrangian}). The target space has then
the maximal symmetry in the class of Lie-algebraic sigma models. Therefore,
to get nontrivial soft theorem in the class of sigma models with 2D target
space, we have to relax the above assumption of Goldstone nature of the
perturbative excitations and take into account more general target spaces
with less symmetries.

In this section we deal with non-linear sigma models with two dimensional
target space equipped with a general metric
\begin{equation}
G\equiv G\left( w,\overline{w}\right) \mathrm{d}w\mathrm{d}\overline{w}
\end{equation}
and the corresponding Lagrangian
\begin{equation}
\mathcal{L}=G\left( \phi ,\overline{\phi }\right)\, \partial \overline{\phi }
\cdot \partial \phi.
\end{equation}
The symmetries of the Lagrangian $\mathcal{L}$ are then in one-to-one
correspondence with the isometries of the target space and can be
represented by the corresponding Killing vectors. For instance, for the
above mentioned Lagrangians $\mathcal{L}_{-}$ (cf. (\ref{CP1 Lagrangian}))
the Killing vectors are
\begin{eqnarray}
L_{-1} &=&\mathrm{i}(1-\alpha \phi )^{2}\partial _{\phi }-\mathrm{i}
(1-\alpha \overline{\phi })^{2}\partial _{\overline{\phi }}\,,  \notag \\[1mm]
L_{0} &=&\frac{1}{2\alpha }\left( 1-\alpha ^{2}\phi ^{2}\right) \partial
_{\phi }+\frac{1}{2\alpha }\left( 1-\alpha ^{2}\overline{\phi }^{2}\right)
\partial _{\overline{\phi }}\,,  \notag \\[1mm]
L_{1} &=&\frac{\mathrm{i}}{4\alpha ^{2}}(1+\alpha \phi )^{2}\partial _{\phi
}-\frac{\mathrm{i}}{4\alpha ^{2}}(1+\alpha \overline{\phi })^{2}\partial _{
\overline{\phi }}\,.  \label{Killing vectors CP1}
\end{eqnarray}
They satisfy the algebra
\begin{equation}
\left[ L_{0},L_{-1}\right] =-L_{-1}\,, \quad \left[ L_{0},L_{1}\right] =L_{1}\,, \quad \left[ L_{-1},L_{1}\right] =-2L_{0} \,. \label{SU(1,1) algebra}
\end{equation}
The above algebra corresponds to generators of $SU\left( 1,1\right) \approx SO(1,2)$
which is the isometry group of the target space $O(1,2)/O(2)$.

At points $w$ for which $G\left( w,\overline{w}\right) \neq 0$ and where 
$G\left( w,\overline{w}\right) $ is analytic in $w,\overline{w}$ we can
perform a weak field expansion writing $\phi \left( x\right) =w+z\left(
x\right) $,\, (i.e. we fix the vacuum expectation value to be $\phi_{\rm vac}\equiv \langle \phi
\rangle =w$) to get the Lagrangian describing small fluctuation near $w$ in
the form
\begin{equation}
\mathcal{L}_{w} \mathcal{=}G\left( w+z,\overline{w}+\overline{z}\right)
\partial \overline{z}\cdot \partial z 
=\partial \overline{z}\cdot \partial z\sum\limits_{n,m}G_{nm}\left( w,%
\overline{w}\right) z^{n}\overline{z}^{m}
\end{equation}
where the couplings are
\begin{equation}
G_{nm}\left( w,\overline{w}\right) =\frac{1}{n!m!}\,\partial _{w}^{n}\,\partial
_{\overline{w}}^{m}\,G\left( w,\overline{w}\right) .
\end{equation}
The corresponding perturbative scattering amplitudes $A_{n_{+}n_{-}}\left(
w\right) $ are then $w$ dependent, this dependence is related to the soft
theorems. In what follows, we denote for short 
\begin{equation}
g\equiv G\left( w,\overline{w}\right) .
\end{equation}
In the next subsection we introduce an example of class of Lagrangians of
this type, which describe gapless non-Goldstone excitations and possess
nontrivial soft theorems. As we will show, the key ingredient for such soft
behavior is the presence of \emph{conformal} Killing vectors on the target
space. These define associated Noether currents to which the elementary
excitations couple. In this sense, this class of Lagrangians 
generalizes the Goldstone boson case (\ref{CP1 Lagrangian}). The
latter appears to be a special element in this class in the limit when the conformal Killing vectors become the ordinary Killing vectors.

\subsection{Introducing the model}\label{subsec31}

Assume the following two-parametric class of Lagrangians for complex scalar $\phi $
\begin{equation}
\mathcal{L}_{\phi }^{K}=\partial \overline{\phi }\cdot \partial \phi \left(
1+\alpha \left( \phi +\overline{\phi }\right) \right) ^{K}  \label{w=1/2a}
\end{equation}
where $K\in \mathbb{R}$ and we suppose $\alpha >0$ with mass dimension $\left[ \alpha \right]
=\left( 2-D\right) /2$ and $D$ is the space-time dimension. It corresponds to
the sigma model on the non-compact manifold with metric
\begin{equation}
G_{\phi \overline{\phi }}=\left( 1+\alpha \left( \phi +\overline{\phi }
\right) \right) ^{K}.
\end{equation}
For general $K$, it is defined for $\mathrm{Re}\phi >-1/2\alpha $. In order
to discuss the symmetries of the target space, it is useful to make a change
of the coordinates 
\begin{equation}
\phi =\psi -\frac{1}{2\alpha },
\label{310}
\end{equation}
which allows us to rewrite the Lagragian and metric as
\begin{equation}
\mathcal{L}_{\psi }^{K}=~\alpha ^{K}\partial \overline{\psi }\cdot \partial
\psi \left( \psi +\overline{\psi }\right)^{K},\qquad G=\alpha^{K}\left( \psi +
\overline{\psi }\right) ^{K}\mathrm{d}\psi \mathrm{d}\overline{\psi }.
\label{L^K_psi}
\end{equation}
The original Lagrangian $\mathcal{L}_{\phi }^{K}$ then describes the
fluctuation around $\psi =\overline{\psi }=(2\alpha)^{-1} $. This particular
choice can be done without loss of generality, since the weak field
expansion around any other point $w$, $\overline{w}$ leads after appropriate
rescaling of the fields\footnote{%
Such a rescaling is necessary in order to get canonical normalization of the
kinetic term.} formally just to $w$, $\overline{w}$ dependent redefinition
of the coupling $\alpha$,
\begin{equation}
\alpha \rightarrow \alpha ^{-K/2}\left( w+\overline{w}\right) ^{-(K+2)/2}.
\label{alpha redefinition}
\end{equation}
The Lagrangian $\mathcal{L}_{\psi }^{K}$ is manifestly invariant with
respect to the infinitesimal shift transformation
\begin{equation}
\delta \psi =\frac{\mathrm{i}}{2\alpha },~~~\delta \overline{\psi }=-\frac{
\mathrm{i}}{2\alpha }
\end{equation}
i.e. there is a Killing vector for the metric $L_{-1}=\mathrm{i}\partial _{\psi }-\mathrm{i}\partial _{\overline{\psi }}$. There are other transformations, under which the Lagrangian transforms in a simple way. Under the infinitesimal scaling $\delta \psi =\psi, \delta \overline{\psi }=\overline{\psi}$, the Lagrangian transforms as
\begin{equation}
\delta \mathcal{L}_{\psi }^{K}=\left( K+2\right) \mathcal{L}_{\psi }^{K}.
\label{delta(-1)L}
\end{equation}
The generator of this transformation on the target space, 
\begin{equation}
L_{0}=\psi \partial _{\psi }+\overline{\psi }\partial _{\overline{\psi }},
\label{delta(0)L}
\end{equation}
is a \emph{conformal} Killing vector for the metric, since the Lie
derivative of the metric is $\mathcal{L}_{L_{0}}G=(K{+}2)G$. Under inversion $I$ defined as
\begin{equation}
\psi \overset{I}{\rightarrow }\frac{1}{\psi },~~~\overline{\psi }\overset{I}{
\rightarrow }\frac{1}{\overline{\psi }}
\end{equation}
we have $\mathcal{L}_{\psi }^{K}\overset{I}{\rightarrow }\left( \psi \overline{\psi }
\right) ^{-K-2}\mathcal{L}_{\psi }^{K}$, i.e. $I$ is a discrete conformal transformation of the target space metric.
Another infinitesimal conformal transformation can be generated by the
combination $I\circ L_{-1}\circ I\equiv L_{1}$, i.e. we get a conformal
Killing vector
\begin{equation}
L_{1}=\mathrm{i}\psi ^{2}\partial _{\psi }-\mathrm{i}\overline{\psi }
^{2}\partial _{\overline{\psi }},
\end{equation}
or in physical notation $\delta \psi =\mathrm{i}\alpha \psi ^{2}, \delta \overline{\psi }=-\mathrm{
i}\alpha \overline{\psi }^{2}$. The Lagrangian transforms under this transformation as
\begin{equation}
\delta \mathcal{L}_{\psi }^{K}=\alpha \left( K+2\right) \left( \psi -
\overline{\psi }\right) \mathcal{L}_{\psi }^{K},  \label{delta(1)L}
\end{equation}
and for the metric we get $\mathcal{L}_{L_{1}}G=\alpha \left( K{+}2\right) \left( \psi {-}\overline{\psi }\right) G$. 
The algebra of the above (conformal) Killing vectors is (cf. (\ref{SU(1,1) algebra}))
\begin{equation}
\left[ L_{0},L_{-1}\right] =-L_{-1}, \quad
\left[ L_{0},L_{1}\right] = L_{1}, \quad
\left[ L_{-1},L_{1}\right] =-2L_{0},
\end{equation}
i.e. it corresponds to $SU\left( 1,1\right) \approx SO(1,2)$. \ Note that
with respect to the above transformations, the metric is invariant for 
$$K=-2\,.$$
For this special value of $K$, the Lagrangian have to be just the
non-compact variant of the $\mathbb{C}P^{1}$. This can be verified directly as follows. 

Let us define new coordinates on the field manifold
\begin{equation}
\psi =\frac{1}{2\alpha }\frac{1+\alpha z}{1-\alpha z},\qquad\overline{\psi }=
\frac{1}{2\alpha }\frac{1+\alpha \overline{z}}{1-\alpha \overline{z}}\,.
\end{equation}
Then the original coordinates in terms of $z$, $\overline{z}$, are 
\begin{equation}
\phi =\frac{z}{1-\alpha z},~~~~\overline{\phi }=\frac{\overline{z}}{1-\alpha 
\overline{z}}\,.
\end{equation}
In the coordinates $z,\overline{z}$, the Lagragian takes the form
\begin{equation}
\mathcal{L}_{z}^{K}\equiv \frac{\partial \overline{z}\cdot \partial z}{\left[
\left( 1-\alpha z\right) \left( 1-\alpha \overline{z}\right) \right] ^{K+2}}
\left( 1-\overline{z}z\right) ^{K}\,,
\end{equation}
and for $K=-2$ we get
\begin{equation*}
\mathcal{L}_{z}^{K{=}-2}= \frac{\partial 
\overline{z}\cdot \partial z}{\left( 1-\alpha ^{2}\overline{z}z\right) ^{2}}\,,
\end{equation*}
as claimed. In these coordinates, the vectors $L_{\pm 1,0}$ are given by (\ref{Killing vectors CP1}) as expected. 

Another interesting case is for $K=-1$,
\begin{equation}
\mathcal{L}_{z}^{K{=}-1}= \frac{\partial 
\overline{z}\cdot \partial z}{\left( 1-\alpha z\right) \left( 1-\alpha 
\overline{z}\right) \left( 1-\alpha ^{2}\overline{z}z\right) }\,,
\end{equation}
corresponding to the Lie-algebraic sigma model with the inverse metric 
\begin{eqnarray}
G^{z\overline{z}}\partial _{z}\partial _{\overline{z}} &=&\left( 1-\alpha
z\right) \left( 1-\alpha \overline{z}\right) \left( 1-\alpha ^{2}\overline{z}%
z\right) \partial _{z}\partial _{\overline{z}}  \notag \\[2mm]
&=&l_{-1}\overline{l}_{-1}-\alpha l_{-1}\overline{l}_{0}-\alpha l_{0}%
\overline{l}_{-1}+\alpha ^{3}l_{1}\overline{l}_{0}+\alpha ^{3}l_{0}\overline{%
l}_{1}-\alpha ^{4}l_{1}\overline{l}_{1}  \label{Lie-algebraic model}
\end{eqnarray}
where $l_{n}=-z^{n+1}\partial _{z}, \overline{l}_{n}=-\overline{z}^{n+1}\partial
_{\overline{z}}$ for $n=0,\pm 1$, are generators of $SL\left( 2,
\mathbb{C}\right) \times SL\left( 2,\mathbb{C}
\right) $. Since
\begin{equation}
\phi =z+O\left( z^{2}\right) ,~~~\phi =\overline{z}+O\left( \overline{z}%
^{2}\right) 
\end{equation}
is an admissible reparameterization of the fields, the amplitudes derived
form the Lagrangian $\mathcal{L}_{z}^{K}$ are the same as those derived from
the original Lagrangian $\mathcal{L}_{\phi }^{K}$. Note that for the 
above-mentioned limits $K\rightarrow -2,\, -1$, also $\mathcal{L}_{\phi }^{K}$ can be
interpreted as a Lie-algebraic sigma model with the inverse metric
\begin{eqnarray}
&&G^{\phi \overline{\phi }}\partial _{\phi }\partial _{\overline{\phi }}\,
\overset{K\rightarrow -2}{=}\ell _{-1}\overline{\ell }_{-1}
+2\alpha \ell
_{-1}\overline{\ell }_{-0}+2\alpha \,\ell _{0}\overline{\ell }_{-1}
\notag \\[1mm]
&&\phantom\qquad\qquad\,\,\,\,\, +2\alpha
^{2}\ell _{0}\overline{\ell }_{0}+\alpha ^{2}\ell _{1}\overline{\ell }
_{-1}+\alpha ^{2}\ell _{-1}\overline{\ell }_{1}\,, \\[1mm]
&&G^{\phi \overline{\phi }}\partial _{\phi }\partial _{\overline{\phi }}\,
\overset{K\rightarrow -1}{=} \,\ell _{-1}\overline{\ell }_{-1}+\alpha \ell _{-1}
\overline{\ell }_{-0}+\alpha \ell _{0}\overline{\ell }_{-1}
\end{eqnarray}
where $\ell _{i}$ and $\overline{\ell }_{i}$ are now given by (\ref{l_i(phi)}). Note also, that the $U\left( 1\right) $ transformation of $\mathcal{L}_{z}^{K}$, namely $\delta z=\mathrm{i}z, \delta \overline{z}=-\mathrm{i}\overline{z}$ can be reformulated in terms of original coordinates as 
\begin{equation}
\delta \phi =\mathrm{i}\phi \left( 1+\alpha \phi \right) ,~~~~~\delta 
\overline{\phi }=-\mathrm{i}\overline{\phi }\left( 1+\alpha \overline{\phi }\right) .
\end{equation}
Under this transformation
\begin{equation}
\delta \mathcal{L}_{\phi }=\mathrm{i}\alpha \left( K+2\right) \left( \phi -
\overline{\phi }\right) \mathcal{L}_{\phi }\,.
\end{equation}
For $K=-2$ this is a $U\left( 1\right) $ charge symmetry, which is
responsible for the Adler zero for the scattering amplitudes in this special
case, as discussed above. Note, however, that this symmetry is not manifest
in the original coordinates $\phi $, $\overline{\phi }$.

To summarize, the class of Lagrangians $\mathcal{L}_{\phi }^{K}$ is rich
enough to encompass two Lie-algebraical sigma models, namely the $%
O(1,2)/O(2) $ model and the model (\ref{Lie-algebraic model}) as special
cases. As we will prove in the next subsection, the symmetry properties of
this class of sigma models allows us to obtain a nontrivial soft theorem,
which substantially differs from that for the Goldstone bosons and
represents another generalization of the Adler zero.

\subsection{The soft theorem}\label{subsec32}

As usual, the key ingredient for the analysis of the soft theorem is the
identification of the currents to which the one-particle states representing
the elementary excitations of the model couple. In our case, these currents
are associated with the Noether currents corresponding to the Killing vector 
$L_{-1}$ \ and the conformal Killing vector $L_{0}$ discussed above.
Denoting them as $J_{-1}$ and $J_{0}$ respectively, we can construct the
following linear combinations 
\begin{eqnarray}
J &\equiv &J_{0}+\mathrm{i}J_{-1}=\frac{1}{\alpha }\partial \phi +K\phi
\partial \phi +\left( K+1\right) \overline{\phi }\partial \phi +\phi
\partial \overline{\phi }+O\left( \phi ^{3},\ldots \right)  \notag \\
\overline{J} &\equiv &J_{0}-\mathrm{i}J_{-1}=\frac{1}{\alpha }\partial 
\overline{\phi }+K\overline{\phi }\partial \overline{\phi }+\left(
K+1\right) \phi \partial \overline{\phi }+\overline{\phi }\partial \phi
+O\left( \phi ^{3},\ldots \right) .  \label{J current}
\end{eqnarray}
At tree level therefore these currents satisfy
\begin{equation}
\langle 0|J\left( 0\right) |p^{+}\rangle =-\frac{\mathrm{i}}{\alpha }
p,~~~\langle 0|\overline{J}\left( 0\right) |p^{-}\rangle =-\frac{\mathrm{i}}{\alpha }p,
\end{equation}
where $|p^{-}\rangle $ and $|p^{+}\rangle $ are one-particle states
corresponding to elementary excitations of the fields $\phi $ and $\overline{\phi }$ respectively around the classical ground state $\phi =\overline{\phi }=0.$ According to (\ref{delta(0)L}) and (\ref{delta(-1)L}), these currents
satisfy the balance equations of the form
\begin{equation}
\partial \cdot J=\partial \cdot \overline{J}=\left( K+2\right) \mathcal{L}_{\phi }.
\end{equation}
This implies the following identities for the matrix elements of the
corresponding operators between ground state and the $n^{+}+n^{-}$ particle $\phi $ and $\overline{\phi }$ in the final state 
\begin{eqnarray}
&&\mathrm{i}\lim_{p\rightarrow 0}p_{\mu }\langle 1^{+},\ldots
,n^{+},1^{-},\ldots ,n^{-}|J^{\mu }\left( 0\right) |0\rangle  \notag \\
&&\hspace{2cm} =\left( K{+}2\right) \int \mathrm{d}^{D}x\langle 1^{+},\ldots
,n^{+},1^{-},\ldots ,n^{-}|\mathcal{L}_{\phi }\left( x\right) |0\rangle .
\label{p.J_matrix_element}
\end{eqnarray}
The matrix element on the left-hand side has a one-particle pole for $p^{2}\rightarrow 0$. At this pole, it factorizes according to 
\begin{eqnarray}
&&\langle 1^{+},\ldots ,n^{+},1^{-},\ldots ,n^{-}|J^{\mu }\left( 0\right)
|0\rangle \label{matrix element of current J}\\ 
&&\hspace{1.5cm}\overset{p^{2}\rightarrow 0}{=}  
\, \langle 0|J^{\mu }\left( 0\right) |p^{+}\rangle \,\frac{\mathrm{i}}{p^{2}}\,
\mathrm{i}A_{n^{+}+1,n^{-}}\left( p^{+},1^{+},\ldots ,n^{-}\right) +{\cal{R}}^{\mu
} \left( p^{+},1^{+},\ldots ,n^{-}\right)  \notag 
\end{eqnarray}
where $A_{n^{+}+n^{-}+1}\left( p^{+},1^{+},\ldots ,n^{-}\right) $ is the
amplitude with an extra $\overline{\phi }\left( p\right) $ particle in the
out state.   The {\em regular} at $p^{2}\rightarrow 0$ term in 
(\ref{matrix element of current J}) is denoted by ${\cal{R}}^{\mu }$. The
left-hand side of (\ref{p.J_matrix_element}) gives then 
\begin{equation}
\frac{\mathrm{i}}{\alpha }\,\lim_{p\rightarrow 0}\,\mathrm{i}A_{n^{+}+1,n^{-}}
\left( p^{+},1^{+},\ldots ,n^{-}\right) +\lim_{p\rightarrow 0}\,\mathrm{i}p_{\mu
}{\cal{R}}^{\mu }\left( p^{+},1^{+},\ldots ,n^{-}\right) .
\end{equation}
Naively, the second term vanishes in the soft limit. However, in fact this is {\em not}
the case, since ${\cal{R}}^{\mu }$ can be singular for $p\rightarrow 0$. The
relevant part of the ``remnant'' ${\cal{R}}_{\rm pole}^{\mu }$ comes from the insertion of the
current into external lines of the Feynman graphs. That means we have to
take into account the quadratic vertices of the current $J$ (cf. (\ref{J
current})). Explicitly,
\begin{eqnarray}
{\cal{R}}_{\rm pole}^{\mu } &=&-\mathrm{i}K\sum\limits_{i=1}^{n^{-}}p^{\mu }\frac{
\mathrm{i}}{2\left( p\cdot k_{i}\right) }\mathrm{i}A_{n^{+},n^{-}}\left(
1^{+},\ldots ,n^{+},1^{-},\ldots ,i^{+},\ldots ,n^{-}\right)  \notag \\
&&-\mathrm{i}\left( K+1\right) \sum\limits_{i=1}^{n^{+}}\mathrm{i}\frac{
p^{\mu }+k_{i}^{\mu }}{2\left( p\cdot k_{i}\right) }\mathrm{i}
A_{n^{+},n^{-}}\left( 1^{+},\ldots ,n^{+},1^{-},\ldots ,n^{-}\right)  \notag
\\
&&+\mathrm{i}\left( K+1\right) \sum\limits_{i=1}^{n^{-}}\mathrm{i}\frac{
k_{i}^{\mu }}{2\left( p\cdot k_{i}\right) }\mathrm{i}A_{n^{+},n^{-}}\left(
1^{+},\ldots ,n^{+},1^{-},\ldots ,n^{-}\right)  \notag \\
&&-\mathrm{i}\sum\limits_{i=1}^{n^{-}}\mathrm{i}\frac{p^{\mu }+k_{i}^{\mu }}{
2\left( p\cdot k_{i}\right) }\mathrm{i}A_{n^{+},n^{-}}\left( 1^{+},\ldots
,n^{+},1^{-},\ldots ,n^{-}\right)  \notag \\
&&+\mathrm{i}\sum\limits_{i=1}^{n^{+}}\mathrm{i}\frac{k_{i}^{\mu }}{2\left(
p\cdot k_{i}\right) }\mathrm{i}A_{n^{+},n^{-}}\left( 1^{+},\ldots
,n^{+},1^{-},\ldots ,n^{-}\right) ,
\end{eqnarray}
where the first line corresponds to the insertion of $K\phi \partial \phi $
into the external $\phi$-line, the next two lines represent the insertion of 
$\left( K+1\right) \overline{\phi }\partial \phi $ into the external $\overline{
\phi }~$\ and $\phi $, respectively, while the last two rows stem from
insertion of $\phi \partial \overline{\phi }$ into $\phi $ and $\overline{
\phi }$ lines, respectively. Therefore,
\begin{equation}
\lim_{p\rightarrow 0}\mathrm{i}p_{\mu }{\cal{R}}_{\rm pole}^{\mu }=\mathrm{i}\frac{K}{2}
\left( n^{+}-n^{-}\right) \mathrm{i}A_{n^{+},n^{-}}\left( 1^{+},\ldots
,n^{+},1^{-},\ldots ,n^{-}\right).
\end{equation}
The right-hand side of (\ref{matrix element of current J}) gives at tree level
\begin{eqnarray}
&&\left( K{+}2\right) \int \mathrm{d}^{D}x\langle 1^{+},\ldots
,n^{+},1^{-},\ldots ,n^{-}|\mathcal{L}_{\phi }\left( x\right) |0\rangle 
\label{lagrangian insertion}\\
&&\hspace{1.5cm} =  \frac{1}{2}\left( K{+}2\right) (2-n^{+}-n^{-})A_{n^{+},n^{-}}\left(
1^{+},\ldots ,n^{+},1^{-},\ldots ,n^{-}\right) .\notag 
\end{eqnarray}
The proof is as follows. Let us first consider the insertion of the kinetic term 
\begin{equation}
\int \mathrm{d}^{D}x\mathcal{L}_{\phi }^{kin}\left( x\right) =\int \mathrm{d}
^{D}x\partial \overline{\phi }\cdot \partial \phi
\end{equation}
with (outgoing) momentum $p\rightarrow 0$ into the external $\overline{\phi }
$ line with (outgoing) momentum $k^{+}$. This gives
\begin{eqnarray}
&&\lim_{p\rightarrow 0}\mathrm{i}k\cdot \left( -\mathrm{i}\right) (p+k)\frac{
\mathrm{i}}{(p+k)^{2}}\mathrm{i}A_{n^{+},n^{-}}\left( 1^{+},\ldots
,k^{+},\ldots ,n^{+},1^{-},\ldots ,n^{-}\right) \notag\\
&&\hspace{2cm} =\frac{\mathrm{i}}{2}\mathrm{i}A_{n^{+},n^{-}}\left( 1^{+},\ldots
,k^{+},\ldots ,n^{+},1^{-},\ldots ,n^{-}\right)
\end{eqnarray}
and a similar expression for the $\phi $-line with the momentum $k^{-}$. Therefore,
\begin{eqnarray}
&&\left( K+2\right) \int \mathrm{d}^{D}x\langle 1^{+},\ldots
,n^{+},1^{-},\ldots ,n^{-}|\mathcal{L}_{\phi }^{kin}\left( x\right)
|0\rangle ^{\rm external}  \notag \\
&&\hspace{1.5cm}=\frac{\mathrm{i}}{2}\left( n^{+}+n^{-}\right) \left( K+2\right) 
\mathrm{i}A_{n^{+},n^{-}}\left( 1^{+},\ldots ,n^{+},1^{-},\ldots
,n^{-}\right) .
\end{eqnarray}
Inserting the kinetic term into the internal line of a general tree-level graph $\Gamma $ gives schematically
\begin{eqnarray}
&&\lim_{p\rightarrow 0}\mathrm{i}A_{L}\left( \ldots ,\left( k+p\right)
^{+},\ldots \right) \frac{\mathrm{i}}{\left( k+p\right) ^{2}}\left( -\mathrm{
i}\right) \left( k+p\right) \cdot \left( -\mathrm{i}\right) \left( -k\right) 
\frac{\mathrm{i}}{k^{2}}\mathrm{i}A_{R}\left( \ldots ,-k^{-},\ldots \right) 
\notag\\
&&\hspace{5cm}=\mathrm{i}\times\mathrm{i}A_{L}\left( \ldots ,k^{+},\ldots \right) \frac{
\mathrm{i}}{k^{2}}\mathrm{i}A_{R}\left( \ldots ,k^{-},\ldots \right),
\end{eqnarray}
and therefore this operation reproduces the original graph up to a factor.
Thus,
\begin{eqnarray}
&&\left( K+2\right) \int \mathrm{d}^{D}x\langle 1^{+},\ldots
,n^{+},1^{-},\ldots ,n^{-}|\mathcal{L}_{\phi }^{kin}\left( x\right)
|0\rangle _{\Gamma }^{\rm internal}  \notag \\
&&\hspace{3cm} =\mathrm{i}I\left( K+2\right)  \mathrm{i}A_{n^{+},n^{-}}\left(
1^{+},\ldots ,n^{+},1^{-},\ldots ,n^{-}\right) _{\Gamma }
\end{eqnarray}
where $I$ is the number of internal lines of a given graph $\Gamma $. Finally, the
insertion of the interaction part of the Lagrangian gives, using the Dirac
interaction picture,
\begin{eqnarray}
&&\int \mathrm{d}^{D}x\langle 1^{+},\ldots ,n^{+},1^{-},\ldots ,n^{-}|
\mathcal{L}_{\phi }^{int}\left( x\right) |0\rangle  \notag\\
&&\hspace{1.5cm}=\left\langle 1^{+},\ldots ,n^{-}\left|T\int \mathrm{d}^{D}x\mathcal{L}_{\phi
}^{int}\left( x\right) _{D}\exp \left( \mathrm{i}\int \mathrm{d}^{D}x
\mathcal{L}_{\phi }^{int}\left( x\right) _{D}\right) \right|0\right\rangle  \notag \\
&&\hspace{1.5cm}=\sum\limits_{n=1}^{\infty }\left( -\mathrm{i}n\right) \frac{\mathrm{i}^{n}
}{n!}\left\langle 1^{+},\ldots ,n^{-}\left|T\left( \int \mathrm{d}^{D}x\mathcal{L}
_{\phi }^{int}\left( x\right) _{D}\right) ^{n}\right|0\right\rangle  \notag,
\end{eqnarray}
and thus for a contribution of the graph $\Gamma $ with $V$ vertices
\begin{eqnarray}
&&\left( K+2\right) \int \mathrm{d}^{D}x\langle 1^{+},\ldots
,n^{+},1^{-},\ldots ,n^{-}|\mathcal{L}_{\phi }^{int}\left( x\right)
|0\rangle _{\Gamma }  \notag \\[2mm]
&&\hspace{2cm} =-\mathrm{i}V\left( K+2\right)  \mathrm{i}A_{n^{+},n^{-}}\left(
1^{+},\ldots ,n^{+},1^{-},\ldots ,n^{-}\right) _{\Gamma }
\end{eqnarray}
where $V$ is number of vertices of the graph $\Gamma $. Assembling together we get
\begin{eqnarray}
&&\left( K+2\right) \int \mathrm{d}^{D}x\langle 1^{+},\ldots
,n^{+},1^{-},\ldots ,n^{-}|\mathcal{L}_{\phi }\left( x\right) |0\rangle
_{\Gamma }  \notag \\
&&\hspace{1.5cm}=\left( \frac{\mathrm{i}}{2}\left( n^{+}+n^{-}\right) +\mathrm{i}I-\mathrm{
i}V\right) \left( K+2\right)  \mathrm{i}A_{n^{+},n^{-}}\left(
1^{+},\ldots ,n^{+},1^{-},\ldots ,n^{-}\right) _{\Gamma }  \notag \\
&&\hspace{1.5cm}=\frac{1}{2}\left( K+2\right) \left( 2-\left( n^{+}+n^{-}\right) \right)
A_{n^{+},n^{-}}\left( 1^{+},\ldots ,n^{+},1^{-},\ldots ,n^{-}\right)
_{\Gamma }\,.
\end{eqnarray}
Using now the formula for tree graphs, $$I=V-1\,,$$ we arrive at Eq. (\ref{lagrangian insertion}). Putting all the ingredients together, as a result we get the following
tree-level soft theorem,
\begin{eqnarray}
&&\lim_{p^+\rightarrow 0}A_{n^{+}+1,n^{-}}\left( p^{+},1^{+},\dots,n^{+},1^{-},\ldots ,n^{-}\right)  \label{soft_theorem}\\
&&\hspace{1cm}= \frac{1}{2}\alpha \left( -K\left( n^{+}-n^{-}\right) +\left( K+2\right)
(2-n^{+}-n^{-})\right) A_{n^{+},n^{-}}\left( 1^{+},\ldots
,n^{+},1^{-},\ldots ,n^{-}\right) , \notag
\end{eqnarray}
and similarly for $p^{-}\rightarrow 0$, with the exchange $%
n^{+}\leftrightarrow n^{-}$. For the maximal symmetry limit $K\rightarrow -2$
we recover the Adler zero (note that in this case, only the amplitudes with 
$n^{+}=n^{-}$ are nonzero due to the $U\left( 1\right) $ symmetry, which is
manifest in the $z$, $\overline{z}$ coordinates). For the second
Lie-algebraic sigma model limit with $K=-1$ we recover the formula (\ref{soft_theorem2}) mentioned in the Introduction. An explicit example and verification of the validity of the soft theorem for 5pt amplitudes can be found in Appendix \ref{appB}.

The soft theorem (\ref{soft_theorem}) can be now used to reconstruct the
tree-level $S$-matrix using the soft BCFW recursion described in the
previous section. The seed amplitudes in this case are
\begin{eqnarray*}
A_{4,0}\left( 1^{+},2^{+},3^{+},4^{+}\right) &=&A_{0,4}\left(
1^{-},2^{-},3^{-},4^{-}\right) = 0, \\
A_{3,1}\left( 1^{-},2^{+},3^{+},4^{+}\right) &=&A_{1,3}\left(
1^{+},2^{-},3^{-},4^{-}\right) =0, \\
A_{2,2}\left( 1^{+},2^{+},3^{-},4^{-}\right) &=&K\alpha ^{2}\left(
1+2\right) \cdot \left( 3+4\right).
\end{eqnarray*}
Combined with the soft theorem, they define the theory uniquely. 

Let us finally briefly discuss the dependence of the tree amplitudes on the point $w$, $\overline{w}$ around which the weak field expansion of the
Lagrangian $\mathcal{L}_{\psi }^{K}$ is performed. As was mentioned in
Subsect. \ref{subsec31}, the only effect of the choice of $w$ different form 
$\psi =\overline{\psi }=1/2\alpha $ is the replacement 
\begin{equation}
\alpha \rightarrow
\alpha ^{-K/2}\left( w+\overline{w}\right) ^{-(K+2)/2}\,.
\end{equation}
By a straightforward dimensional analysis we find that $ A_{n^{+},n^{-}}\propto \alpha ^{\left( n^{+}+n^{-}-2\right)}$ and thus the dependence on $w$ is simply
\begin{equation}
A_{n^{+},n^{-}}\left( w\right) \propto \left( w+\overline{w}\right) ^{\frac{1
}{2}(K+2)\left( 2-n^{+}-n^{-}\right) }.
\end{equation}
Therefore, we have the following symmetry relation for $A_{n^{+},n^{-}}
\left( w\right) $ with respect to the (conformal) Killing vectors of the
target space
\begin{eqnarray}
L_{-1}^{w}A_{n^{+},n^{-}}\left( w\right) &\equiv &\mathrm{i}\left( \partial
_{w}-\partial _{\overline{w}}\right) A_{n^{+},n^{-}}\left( w\right) =0, \label{our model symmetries}\\
L_{0}^{w}A_{n^{+},n^{-}}\left( w\right) &\equiv &\left( w\partial _{w}+
\overline{w}\partial _{\overline{w}}\right) A_{n^{+},n^{-}}\left( w\right) =
\frac{1}{2}(K+2)\left( 2-n^{+}-n^{-}\right) A_{n^{+},n^{-}}\left( w\right),\notag\\
L_{1}^{w}A_{n^{+},n^{-}}\left( w\right) &\equiv &\mathrm{i}\left(
w^{2}\partial _{w}-\overline{w}^{2}\partial _{\overline{w}}\right)
A_{n^{+},n^{-}}\left( w\right) =\frac{1}{2}(K+2)\left( 2-n^{+}-n^{-}\right)
\left( w-\overline{w}\right) A_{n^{+},n^{-}}\left( w\right). \notag
\end{eqnarray}

\section{Generalizations}\label{sec4}

In the previous section we have discussed a particular model with the interesting soft theorem of non-Goldstone type. In fact, it was a special case of more general models, for which we can derive similar soft theorems, or rather
\textquotedblleft soft sum rules\textquotedblright\ combining soft limits of
different amplitudes. Again, the key ingredient is the existence of the
appropriate (conformal) Killing vector on the target space, associated with a
current which couples to some superposition of one-particle states. Let us
give two examples of a wider class of such non-linear sigma models.

\subsection{Homogeneous metric}\label{subsec41}

In the first example, we assume the case of just one special conformal
Killing vector. Suppose, that the Lagrangian has the form
\begin{equation}
\mathcal{L}=G\left( \phi ,\overline{\phi }\right) \partial \phi \cdot
\partial \overline{\phi }  \label{homogeneous metric}
\end{equation}
where $G\left( z,\overline{z}\right) $ is homogeneous of degree $K$, i.e.
\begin{equation}
G\left( \lambda \phi ,\lambda \overline{\phi }\right) =\lambda ^{K}G\left(
\phi ,\overline{\phi }\right) .  \label{homogenity}
\end{equation}
The sigma model discussed in the previous section with Lagrangian (\ref
{L^K_psi}) is then a special case. The desired conformal Killing vector is
then
\begin{equation}
\ell _{0}=\phi \partial _{\phi }+\overline{\phi }\partial _{\overline{\phi }}
\end{equation}
which corresponds to the infinitesimal scaling $\delta \phi =\phi, \delta 
\overline{\phi }=\overline{\phi }$. Under this transformation we have
\begin{equation}
\delta \mathcal{L}=\left( K+2\right) \mathcal{L},~~~~L_{\ell _{0}}G=\left(
K+2\right) G.
\end{equation}
The weak field expansion $\phi \left( x\right) =w+z\left( x\right) $ around
the vacuum point $\{w,\overline{w}\} $ gives a family of theories with the Lagrangians
\begin{equation}
\mathcal{L}_{w} = G\left( z+w,\overline{z}+\overline{w}\right) \partial
z\cdot \partial \overline{z}  
=\partial z\cdot \partial \overline{z}\left[ g+z\partial _{w}g+\overline{z}
\partial _{\overline{w}}g+\ldots \right]
\end{equation}
where we denoted $G\left( w,\overline{w}\right) \equiv g$ for short. The
infinitesimal scaling is realized nonlinearly
\begin{equation}
\delta z=z+w,~~~\delta \overline{z}=\overline{z}+\overline{w}\,,
\end{equation}
and the corresponding Noether current reads
\begin{eqnarray}
J &=&G\left( z+w,\overline{z}+\overline{w}\right) \left[ \left( \overline{z}+
\overline{w}\right) \partial z+\left( z+w\right) \partial \overline{z}\right]\\
&=&g\overline{w}\partial z+gw\partial \overline{z}+(g+\overline{w}\partial _{
\overline{w}}g)\overline{z}\partial z+(g+w\partial _{w}g)z\partial \overline{
z}  +\overline{w}\partial _{w}gz\partial z+w\partial _{\overline{w}}g\overline{
z}\partial \overline{z}+O\left( z^{3}\right). \notag
\end{eqnarray}
This current is not conserved, but its divergence equation is known,
\begin{equation}
\partial \cdot J=\left( K+2\right) \mathcal{L}_{w}.
\end{equation}
It also does not couple to a single particle state corresponding to the fields 
$z$ and $\overline{z}$, but rather to their linear combination $w|p^{+}\rangle +\overline{w}|p^{-}\rangle $. Nevertheless we can readily repeat the general method from the previous section to prove the following
``soft sum rule'' in the form of the soft limit
of a linear combination of amplitudes with different ``flavors'' of the soft
particle,
\begin{eqnarray}
&&\lim_{p\rightarrow 0}g^{1/2}\left( \overline{w}
A_{n^{+}+1,n^{-}}+wA_{n^{+},n^{-}+1}\right) \label{soft sum rule homogenity}\\
&&\hspace{1.5cm}=\frac{1}{2}\left[ \left( n^{+}-n^{-}\right) \left( g^{-1}w\partial
_{w}g-g^{-1}\overline{w}\partial _{\overline{w}}g\right) +(K+2)\left(
2-n^{+}-n^{-}\right) \right] A_{n^{+},n^{-}}.\notag
\end{eqnarray}
It is an easy exercise to show, that the sigma model discussed in the
Sect. \ref{non GB soft} with Lagrangian $\mathcal{L}_{\psi }^{K}$ (see (\ref{L^K_psi})) 
satisfies this relation for $w=\overline{w}=1/2\alpha$ where the parameter $\alpha$ is defined in (\ref{310}).

\subsection{{\boldmath$U(1)$} invariant metric}

In the given example, the target space possesses a true Killing vector with the
desired properties. Suppose, that the Lagrangian has the form
\begin{equation}
\mathcal{L}=G\left( \phi \overline{\phi }\right) \partial \phi \cdot
\partial \overline{\phi }  \label{U(1) symmetric metric}
\end{equation}
where $G\left( \cdot \right) $ is a real function, i.e. $\mathcal{L}$ is
invariant with respect to the $U\left( 1\right) $ rotations
\begin{equation*}
\phi \rightarrow \mathrm{e}^{\mathrm{i}\alpha }\phi ,~~~\overline{\phi }
\rightarrow \mathrm{e}^{-\mathrm{i}\alpha }\overline{\phi }\,.
\end{equation*}
In particular,  a class of the Lie-algebraic models that has been  previously considered in \cite{Sheu:2024sxf} is just a small subclass of  (\ref{U(1) symmetric metric}). As before, let us define the shifted theory by weak field expansion around $\{
w, \overline{w}\}$,
\begin{equation}
\mathcal{L}_{w}=G\left( \left( z+w\right) \left( \overline{z}+\overline{w}
\right) \right) \partial z\cdot \partial \overline{z}
\label{411}
\end{equation}
which is invariant with respect to the infinitesimal generalized shift
(non-linear realization of the original $U\left( 1\right) $ rotation)
\begin{equation}
\delta z=\mathrm{i}\left( z+w\right) ,~~~\delta z=-\mathrm{i}\left( 
\overline{z}+\overline{w}\right) .
\end{equation}
The corresponding conserved Noether current reads
\begin{eqnarray}
J &=&-\mathrm{i}G\left( \left( z+w\right) \left( \overline{z}+\overline{w}
\right) \right) \left[ \mathrm{i}\left( z+w\right) \partial \overline{z}-
\mathrm{i}\left( \overline{z}+\overline{w}\right) \partial z\right]\\
&=&g\left( w\partial \overline{z}-\overline{w}\partial z\right) +\left( g+w
\overline{w}g^{\prime }\right) \left( z\partial \overline{z}-\overline{z}
\partial z\right) +g^{\prime }\left( w^{2}\overline{z}\partial \overline{z}-
\overline{w}^{2}z\partial z\right) +O\left( z^{3}\right),  \notag
\end{eqnarray}
with $\partial \cdot J =0$ where as above we abbreviated
\begin{equation}
g\equiv G\left( w\overline{w}\right) ,~~~g^{\prime }\equiv G^{\prime }\left(
w\overline{w}\right) .
\end{equation}
Now the current couples to the combination $(w|p^{+}\rangle -\overline{w}|p^{-}\rangle)$
and therefore, using the method of the previous section, we
derive a ``soft sum rule'' in the form
\begin{equation}
\lim_{p\rightarrow 0}g^{1/2}\left( wA_{n^{+},n^{-}+1}-\overline{w}
A_{n^{+}+1,n^{-}}\right) =\left( n^{+}-n^{-}\right) \left( 1+w\overline{w}
g^{-1}g^{\prime }\right) A_{n^{+},n^{-}}\,.  \label{soft sum rule U(1)}
\end{equation}
For generic choices of vacua $\{w,\,\overline{w}\}$ the apparent $U(1)$ symmetry of the Lagrangian (\ref{U(1) symmetric metric}) is spontaneously broken and 
charge conservation is realized non-linearly. This means that 
the soft sum rule in (\ref{soft sum rule U(1)}) is informative for any values of $n^+,\,n^-$. However, there may exist special points on the target space corresponding to 
the charge-conserving vacua. An obvious example is $w=\overline{w} =0$ in (\ref{411}).
In this case Eq. (\ref{soft sum rule U(1)}) degenerates into the trivial charge conservation condition
\begin{equation}
    (n^+-n^-)A_{n^+,n^-}=0
\end{equation}
and thus only the amplitudes with $n^+=n^-$ (if all lines are outgoing)  do not vanish.

\subsection{General case without symmetries}
\label{43}

The core of our method for deriving the above soft theorems (or soft sum rules)
was the existence of distinguished currents which couple to the
one-particle states. Such currents were associated with some special
transformations, under which the target space metric was invariant or at
least conformally invariant. In all the cases, the balance equation for the currents was simple enough to relate the matrix elements of the divergence of the current back to the amplitudes.

Nevertheless, even if we have no such currents associated with nice
symmetries of the target space at our disposal, there are always two
currents coupled to the one-particle states under which the Lagrangian of
the theory transforms in a simple way. These are the Noether currents
corresponding to the pure shift transformation of the fields. We will see,
that in such a case we can derive a soft theorem, which is in fact a special
case of those derived in \cite{Cheung:2021yog}. Let us consider the general Lagrangian of the form
\begin{equation}
\mathcal{L}=G\left( \phi ,\overline{\phi }\right) \partial \phi \cdot
\partial \overline{\phi }
\end{equation}
Since the general Lagrangian is generally not invariant with respect to the
shift symmetry, the perturbative weak field expansion 
\begin{equation*}
\phi \left( x\right) =w+\psi \left( x\right) ,~~~\overline{\phi }\left(
x\right) =\overline{w}+\overline{\psi }\left( x\right)
\end{equation*}
around fixed constant $w$, $\overline{w}$ for which $g\equiv G\left( w,\overline{w}\right) \neq 0$ 
generates a family of theories with Lagrangians
\begin{align}
\mathcal{L}_{w} &=G\left( \psi +w,\overline{\psi }+\overline{w}\right)
\partial \psi \cdot \partial \overline{\psi }\\
&=\partial \psi \cdot \partial \overline{\psi }\left[ G\left( w,\overline{w}
\right) +\psi \partial _{w}G\left( w,\overline{w}\right) +\overline{\psi }
\partial _{\overline{w}}G\left( w,\overline{w}\right) +\ldots \right]\nonumber
\end{align}
In what follows, we will concentrate to the amplitudes $A_{n^{+}+n^{-}}\left( w\right) $ generated by the Lagrangian $\mathcal{L}_{w}$. It is useful to redefine the fields by means of rescaling
\begin{equation*}
\psi =g^{-1/2}z,~~~\overline{\psi }=g^{-1/2}\overline{z},
\end{equation*}
in order to get canonically normalized kinetic term and the usual Feynman
rule for the perturbative propagator. The Noether currents corresponding to
the pure shift transformation $z\rightarrow z+c,\, \overline{z}\rightarrow \overline{z}+\overline{c}$ are
\begin{align}
J &=g^{-1}G\left( w+g^{-1/2}z,\overline{w}+g^{-1/2}\overline{z}\right)
\partial z=\left[ 1+g^{-3/2}z\partial _{w}g+g^{-3/2}\overline{z}\partial _{%
\overline{w}}g+\ldots \right] \partial z\, ,\nonumber \\
\overline{J} &=g^{-1}G\left( w+g^{-1/2}z,\overline{w}+g^{-1/2}\overline{z}%
\right) \partial \overline{z}=\left[ 1+g^{-3/2}z\partial _{w}g+g^{-3/2}%
\overline{z}\partial _{\overline{w}}g+\ldots \right] \partial \overline{z}\,. \label{currents general}
\end{align}
On shell, these currents satisfy the balance equations which just coincide
with the Euler-Lagrange equations of motion 
\begin{equation}
\partial \cdot J=\frac{\partial \mathcal{L}_{w}}{\partial \overline{z}}
=g^{-1/2}\,\frac{\partial \mathcal{L}_{w}^{int}}{\partial \overline{w}}
\,,~~~~~\partial \cdot \overline{J}=\frac{\partial \mathcal{L}_{w}}{\partial z}
=g^{-1/2}\,\frac{\partial \mathcal{L}_{w}^{int}}{\partial w}\,.  \notag
\end{equation}
At tree level we have then
\begin{equation}
\langle 0|J\left( 0\right) |p^{+}\rangle =-\mathrm{i}p,~~~\langle 0|
\overline{J}\left( 0\right) |p^{-}\rangle =-\mathrm{i}p,
\end{equation}
and therefore, as usual, the tree-level amplitude can be identified as the
residue at the one-particle pole for $p^{2}\rightarrow 0$ of the matrix
element of $J$
\begin{equation}
\left\langle n^{+},n^{-}|{J}\left(0\right) |0\right\rangle =-
\mathrm{i}p\frac{\mathrm{i}}{p^{2}}\mathrm{i}A_{n^{+}+1,n^{-}}+{\cal{R}}\,,
\end{equation}
where ${\cal{R}}$ is a regular part in the limit $p^2\to 0$ and 
\begin{equation}
\mathrm{i}A_{n^{+}+1,n^{-}}=\left\langle p^{+},n^{+},n^{-}|0\right\rangle\, .
\end{equation}
Using now the same consideration as in the previous sections, i.e. using the balance equation for the current in the momentum representation in the form
\begin{equation}
-\lim_{p\rightarrow 0}\mathrm{i}p\cdot \left\langle n^{+},n^{-}|{J}
\left( 0\right) |0\right\rangle =g^{-1/2}\int \mathrm{d}^{D}x\left\langle
n^{+},n^{-}|\partial _{\overline{w}}\mathcal{L}_{w}^{int}\left( x\right)
|0\right\rangle
\end{equation}
and extracting the one-particle pole contribution we arrive at
\begin{equation}
\lim_{p\rightarrow 0}A_{n^{+}+1,n^{-}}=\lim_{p\rightarrow 0}\mathrm{i}p\cdot
R+g^{-1/2}\int \mathrm{d}^{D}x\left\langle n^{+},n^{-}|\partial _{\overline{w
}}\mathcal{L}_{w}^{int}\left( x\right) |0\right\rangle \,.
\end{equation}
As shown in detail in the Appendix \ref{appA}, the contributions of the two terms on the right-hand side combine into the resulting soft theorem of the form
\begin{equation}
\lim_{p\rightarrow 0}A_{n^{+}+1,n^{-}}=g^{-1/2}\left[ -\frac{1}{2}\left(
n^{+}-n^{-}\right) g^{-1}\left( \partial _{\overline{w}}g\right) +\partial _{
\overline{w}}\right] A_{n^{+},n^{-}},  \label{general_soft_theorem+}
\end{equation}
and similarly
\begin{equation}
\lim_{p\rightarrow 0}A_{n^{+},n^{-}+1}=g^{-1/2}\left[ -\frac{1}{2}\left(
n^{-}-n^{+}\right) g^{-1}\left( \partial _{w}g\right) +\partial _{w}\right]
A_{n^{+},n^{-}}.  \label{general_soft_theorem-}
\end{equation}
That means, these general soft theorems probe the dependence of the
perturbative $S-$matrix on the chosen vacuum expectation value 
$\left\langle \phi \right\rangle =w$, $\left\langle \overline{\phi }
\right\rangle =\overline{w}$ around which we perform the weak field
expansion. As we have seen in the previous subsections, provided the target
space has additional symmetries corresponding to generalized shift
transformation of the fields, we can derive soft sum rules, which are in
general independent on (\ref{general_soft_theorem+}) and (\ref{general_soft_theorem-}). 
If this is the case, we can combine them to obtain
additional symmetry relations for the amplitudes. For example, for the
homogeneous metric (\ref{homogeneous metric}), using the soft sum rule (\ref{soft sum rule homogenity})
 we get transformation property of the amplitude 
 $A_{n^{+}+n^{-}}$ with respect to the scaling in the form 
\begin{equation}
DA_{n^{+},n^{-}}=\frac{1}{2}(K+2)\left( 2-n^{+}-n^{-}\right) A_{n^{+},n^{-}},
\end{equation}
where $$D=-L_{0}^{w}-\overline{L}_{0}^{w}=w\partial _{w}+\overline{w}\partial _{
\overline{w}}$$ is the corresponding conformal Killing vector. Similarly, for the $U\left(
1\right) $ invariant metric (\ref{U(1) symmetric metric}), we have with help
of the soft sum rule (\ref{soft sum rule U(1)}) the following transformation
of the amplitude with respect to the $U\left( 1\right) $ rotations 
\begin{equation}
TA_{n_{+}n_{-}}\left( w\right) =\left( n_{+}-n_{-}\right)
A_{n_{+},n_{-}}\left( w\right)
\end{equation}
with the Killing vector
\begin{equation}
T=-L_{-1}^{w}+\overline{L}_{-1}^{w}=w\partial _{w}-\overline{w}\partial _{
\overline{w}}.
\end{equation}
The last example is the model discussed in Section \ref{non GB soft} with
the target space metric 
\begin{equation*}
g=\alpha ^{K}\left( w+\overline{w}\right) ^{K}.
\end{equation*}
It is easily seen that the symmetry relations (\ref{our model symmetries})
can be obtained alternatively by means of combining (\ref{general_soft_theorem+}) 
and (\ref{general_soft_theorem-}) with the soft theorem (\ref{soft_theorem}).

Remarkably, as proved in \cite{Cheung:2021yog} for the first time, the soft
theorems (\ref{general_soft_theorem+}) and (\ref{general_soft_theorem-}) can
be interpreted in a nice geometric way. Namely, let us define the tensor of
the rank $n^{+}+n^{-}$ with components 
\begin{equation}
\Gamma _{\overline{z}\ldots \overline{z}z\ldots z}\equiv g^{\frac{1}{2}
\left( n^{+}+n^{-}\right) }A_{n^{+},n^{-}}\left( w\right) ,
\end{equation}
where $n_{+}$ and $n_{-}$ are the numbers of $\overline{z}$ and $z$ indices,
respectively. Note that the Christoffel symbols corresponding to the
Levi-Civita connection associated with the metric $G\left( w,\overline{w}\right) $ are
\begin{equation}
\Gamma _{\overline{w}\overline{w}}^{w} =\Gamma _{\overline{w}w}^{w}=\Gamma
_{\overline{w}w}^{\overline{w}}=\Gamma _{ww}^{\overline{w}}=0, \quad
\Gamma _{\overline{w}\overline{w}}^{\overline{w}} =g^{-1}\left( \partial _{
\overline{w}}g\right) ,~~~\Gamma _{ww}^{w}=g^{-1}\partial_{w}g\,.
\end{equation}
Then (\ref{general_soft_theorem+}) can be rewritten in the form of the geometric
soft theorem (cf. \cite{Cheung:2021yog})
\begin{equation}
\lim_{p\rightarrow 0}\Gamma _{\overline{w}\overline{z}\ldots \overline{z}
z\ldots z}=\left[ \partial _{\overline{w}}-n^{+}g^{-1}\left( \partial _{
\overline{w}}g\right) \right] \Gamma _{\overline{z}\ldots \overline{z}
z\ldots z}=\nabla _{\overline{w}}\Gamma _{\overline{z}\ldots \overline{z}
z\ldots z}
\label{eq433}
\end{equation}
where $\nabla_{\overline{w}}$ is the covariant derivative. Similarly 
\begin{equation}
\lim_{p\rightarrow 0}\Gamma _{\overline{z}\ldots \overline{z}z\ldots zw}=
\left[ \partial _{w}-n^{-}g^{-1}\left( \partial _{w}g\right) \right] \Gamma
_{\overline{z}\ldots \overline{z}z\ldots z}=\nabla _{w}\Gamma _{\overline{z}
\ldots \overline{z}z\ldots z}\,.
\label{eq434}
\end{equation}
Hence, the soft limits just calculate the covariant derivatives of the
rescaled amplitude $\Gamma _{\overline{z}\ldots \overline{z}z\ldots z}$ with
respect to the vacuum expectation values. The general soft theorems (\ref{general_soft_theorem+}) and (\ref
{general_soft_theorem-}) can be also used in order to reconstruct the
amplitudes using soft recursion. The seed amplitude is now
\begin{eqnarray}
A_{2,2}\left( 1^{+},2^{+},3^{-},4^{-}\right) &=&\frac{1}{g^{2}}\left[
g^{-1}\partial _{w}g~\partial _{\overline{w}}g-\partial _{\overline{w}
}\partial _{w}g\right] \left( 1+2\right) \cdot \left( 3+4\right)  \notag \\
&=&-\frac{1}{g^{2}}R_{\overline{w}w\overline{w}w}\left( 1+2\right) \cdot
\left( 3+4\right) 
\label{general_4pt}
\end{eqnarray}
where 
\begin{equation}
R_{\overline{w}w\overline{w}w}=\partial _{\overline{w}}\partial
_{w}g-g^{-1}\partial _{w}g~\partial _{\overline{w}}g
\label{general_curvature}
\end{equation}
is the target space curvature tensor. The other 4pt amplitudes vanish. The
higher-point amplitudes can be then reconstructed uniquely in terms of $%
A_{4}\left( 1^{+},2^{+},3^{-},4^{-}\right) $ using the soft BCFW recursion
based on the soft theorems (\ref{general_soft_theorem+}) and (\ref%
{general_soft_theorem-}). 

We pause here to add an extra remark regarding the difference between 
our recursion method in Sect. \ref{non GB soft} and the geometric one by Cheung et al. \cite{Cheung:2021yog}.
A prerequisite to the derivation of geometric soft theorems is the knowledge of geometric data on the target space. The right-hand sides in (\ref{eq433}) and  (\ref{eq434}) contain the covariant derivatives (i.e. the Christoffel symbols or derivatives of the metric as a function of $w,\overline{w}$), implying that
we have to know the metric and therefore also the Lagrangian {\em before} we try to define the theory via soft recursion for the amplitudes.  Of course, one could try to prescribe some particular consistent form of the Christoffel symbols without any reference to the Lagrangian. This would fix the right-hand side of the soft theorems. After that one could launch the recursive construction of the amplitudes. To this end, in addition, one would have to fix the seed amplitudes as functions of $w,\overline{w}$. Alas, this step is {\em not} arbitrary  -- it must be in accord with the choice of the Christoffel symbols.
Moreover, the 4pt amplitude is proportional to the curvature and the inverse metric squared as functions of $w,\overline{w}$ (see Eq. (\ref{general_4pt}) and (\ref{general_curvature}) above).

Thus, the information contained in the right-hand sides of the geometric soft theorems plus the seed amplitudes for recursion is equivalent to the {\em a priori} knowledge of the Lagrangian. Then  the recursion is {\em not} needed at all 
and the soft theorems (\ref{general_soft_theorem+}) and  (\ref{general_soft_theorem-}) cannot be used as a completely independent alternative Lagrangian-free definition of the theory. On the other hand, in the case of the soft theorem (\ref{soft_theorem}), the information hidden in it and in the seed amplitude (now the latter is determined  uniquely, up to an overall normalization constant) does not allow one to easily deduce the Lagrangian at first sight. In order to identify the theory from (\ref{soft_theorem}), the recursion is inevitable. 

\subsection{Double soft limit}\label{subsec44}

The results of the previous subsections allow us to discuss also multiple
soft limits. As found in \cite{Cheung:2021yog}, the latter probe further
characteristics of the geometry of the target space, namely the Riemann
curvature. Here we give an elementary derivation of the various versions of
the double soft limits using the methods developed in \cite{Kampf:2013vha, Low:2018acv} and \cite{Cheung:2021yog}
and apply the results to our model discussed in Sect. \ref{non GB soft}.
Similarly as for the single soft limits, in this particular case the double
soft limit does not involve derivatives of the amplitudes, so we can
completely avoid the necessity to know the dependence of the amplitude on
the vacuum expectation value, which makes our model special. To begin with, let us assume the general case of fluctuations around $\phi=w $, $\overline{\phi }=\overline{w}$ 
\begin{eqnarray}
\mathcal{L}_{w} &=&G\left( z+w,\overline{z}+\overline{w}\right) \partial
z\cdot \partial \overline{z}  \notag \\[1mm]
&=&\partial z\cdot \partial \overline{z}\left[ g+z\partial _{w}g+\overline{z}
\partial _{\overline{w}}g+\frac{1}{2}z^{2}\partial _{w}^{2}g+\frac{1}{2}
\overline{z}^{2}\partial _{\overline{w}}^{2}g+z\overline{z}\partial
_{w}\partial _{\overline{w}}g+\ldots \right]
\end{eqnarray}
where as above $g\equiv G\left( w,\overline{w}\right)$. In order to avoid the contributions stemming from the cubic vertices, we can
eliminate them using the following field redefinition which does not change
the $S$-matrix
\begin{equation}
z \rightarrow z-\frac{1}{2}z^{2}g^{-1}\partial _{w}g \,, \quad
\overline{z} \rightarrow \overline{z}-\frac{1}{2}\overline{z}
^{2}g^{-1}\partial _{\overline{w}}g\,.
\end{equation}
In these new fields, the interaction Lagrangian starts with quartic vertices
which simplifies the further analysis. Explicitly we obtain
\begin{eqnarray}
\mathcal{L}_{w} &=&g\partial z\cdot \partial \overline{z}+\partial z\cdot \partial \overline{z}\frac{1}{2g}\left[ \left( \partial
_{w}\partial _{w}g-3\partial _{w}g\partial _{w}g\right) z^{2}+\left(
\partial _{\overline{w}}\partial _{\overline{w}}g-3\partial _{\overline{w}
}g\partial _{\overline{w}}g\right) \overline{z}^{2}\right]\notag  \\
&&+\partial z\cdot \partial \overline{z}\frac{1}{g}\left( g\partial _{
\overline{w}}\partial _{w}g-\partial _{w}g\partial _{\overline{w}}g\right) z
\overline{z}+\ldots\,.
\end{eqnarray}
After further rescaling $z\rightarrow g^{-1/2}z$,\,\,\, $\overline{z}\rightarrow
g^{-1/2}\overline{z}$ we recover the canonical normalization of the kinetic
term and obtain the following momentum-space Feynman rules for 4pt vertices
(here all momenta are treated as outgoing),
\begin{eqnarray}
&&
V_{4}\left( 1^{+}2^{+}3^{+}4^{-}\right) =\frac{1}{g^{3}}\left( \partial _{
\overline{w}}\partial _{\overline{w}}g-3\partial _{\overline{w}}g\partial _{
\overline{w}}g\right) p_{4}^{2}, \nonumber\\[1mm]
&&
V_{4}\left( 1^{-}2^{-}3^{-}4^{+}\right) =\frac{1}{g^{3}}\left( \partial
_{w}\partial _{w}g-3\partial _{w}g\partial _{w}g\right) p_{4}^{2}\,,  \notag\\[1mm]
&&  V_{4}\left( 1^{+}2^{+}3^{-}4^{-}\right) =-\frac{R}{g^{2}}\left(
p_{1}+p_{2}\right) \cdot \left( p_{3}+p_{4}\right)  , \label{4pt vertices}
\end{eqnarray}
where we abbreviated $$R\equiv R_{\overline{w}w\overline{w}w}=\partial _{\overline{w}}\partial
_{w}g-g^{-1}\partial _{w}g~\partial _{\overline{w}}g\,.$$  Let us now assume the amplitude with $n_{+}+1$ positive charge particles
(i.e. with $n_{+}+1$ external $z$ legs) and $n_{-}+1$ negative charge
particles (with $n_{-}+1$ external $\overline{z}$ legs). In order to probe the soft limit of two particles, say $1$ and $2$, we can
proceed as follows. 

Let us start with $(n_{+}+n_{-})$-particle configuration with 
$p_{i}$, $i=3,\ldots n_{+}+n_{-}$, satisfying the on-shell conditions $
p_{i}^{2}=0$ and momentum conservation
\begin{equation}
\sum\limits_{i=3}^{n_{+}+n_{-}+2}p_{i}=0,
\end{equation}
and add two arbitrary on-shell momenta $p_{1}$ and $p_{2}$. \ To get a valid 
$n_{+}+n_{-}+2$ configuration which allows to perform the double soft limit,
we construct an appropriate deformation of the original on-shell momenta $
p_{i}$, $i=1,\ldots ,n_{+}+n_{-}+2$ depending on two parameters $u,v\in
\left\langle 0,1\right\rangle $ 
\begin{equation}
p_{i}\rightarrow \widehat{p}_{i}\left( u,v\right) ,
\end{equation}
in such a way that the soft momenta $\widehat{p}_{1}$ and $\widehat{p}_{2}$
simply scale with these parameters as 
\begin{eqnarray}
\widehat{p}_{1}\left( u,v\right) = up_{1} \,,\quad  
\widehat{p}_{2}\left( u,v\right) =vp_{2}\,,
\end{eqnarray}
i.e. the soft limit is achieved for $u\rightarrow 0$ and/or $v\rightarrow 0$.
The other deformed momenta should satisfy the on-shell condition 
$\widehat{p}_{i}\left( u,v\right) ^{2}=0$, the momentum conservation 
\begin{equation}
\sum\limits_{i=1}^{n_{+}+n_{-}+2}\widehat{p}_{i}\left( u,v\right) =0
\end{equation}
for all $u,v$ and the initial condition $\widehat{p}_{i}\left( 0,0\right)
=p_{i}$. In what follows we will implicitly assume the existence of such a
deformation. The particular realization of the deformation will not be
important for our discussion.

Let us start with the case when the soft particles have opposite charges.
Various soft theorems correspond to different ways of approaching the limit 
$u\rightarrow 0$ and $v\rightarrow 0$ of the deformed amplitude (here we
again abbreviate $p_{i}\rightarrow i$) 
\begin{equation}
A_{+-}(u,v)\equiv A_{n_{+}+1,n_{-}+1}\big(u1^{+},v2^{-},\ldots ,\widehat{i}
(u,v),\ldots \big).
\end{equation}
Independently of the way this limit is performed, the amplitude can be
decomposed picking out the terms which correspond to the attachment of the
4pt vertex with two soft particles to the rest of the amplitude. These terms
contain propagators which are singular in the limit $u,v\rightarrow 0$. The
decomposition is as follows,
\begin{eqnarray}
A_{+-}(u,v) &=&\sum\limits_{i^{+}}V_{4}\left( u1^{+},\widehat{i}^{+},-\left(
u1+v2+\widehat{i}\right) ^{+},v2^{-}\right) \frac{\mathrm{i}}{\left( u1+v2+
\widehat{i}\right) ^{2}}\mathrm{i}\widehat{A}_{n_{+}-1,n_{-}+1}^{(i^{-})} 
\notag \\[1mm]
&&+\sum\limits_{i^{-}}V_{4}\left( u1^{+},\widehat{i}^{-},-\left( u1+v2+
\widehat{i}\right) ^{-},v2^{-}\right) \frac{\mathrm{i}}{\left( u1+v2+
\widehat{i}\right) ^{2}}\mathrm{i}\widehat{A}_{n_{+}+1,n_{-}-1}^{(i^{-})} 
\notag \\[1mm]
&&+\sum\limits_{i^{+}}V_{4}\left( u1^{+},\widehat{i}^{+},-\left( u1+v2+
\widehat{i}\right) ^{-},v2^{-}\right) \frac{\mathrm{i}}{\left( u1+v2+
\widehat{i}\right) ^{2}}\mathrm{i}\widehat{A}_{n_{+},n_{-}}^{(i^{+})}  \notag
\\[1mm]
&&+\sum\limits_{i^{-}}V_{4}\left( u1^{+},-\left( u1+v2+\widehat{i}\right)
^{+},\widehat{i}^{-},v2^{-}\right) \frac{\mathrm{i}}{\left( u1+v2+\widehat{i}
\right) ^{2}}\mathrm{i}\widehat{A}_{n_{+},n_{-}}^{(i^{-})}  \notag \\
&&+\,\mathcal{R}_{+-}\left( u,v\right) .  \notag  \label{A+- decomposition}
\end{eqnarray}
Here $\mathcal{R}_{+-}\left( u,v\right) $ is the remnant (a regular part) which does not
develop pole when $u,v\rightarrow 0$), and
\begin{equation}
\widehat{A}_{n_{+},n_{-}}^{(i^{\pm })}\equiv \widehat{A}_{n_{+},n_{-}}\left(
\ldots ,\left( \widehat{i}+i1+v2\right) ^{\pm },\ldots ,\right)
\end{equation}
corresponds to the off-shell extension of the amplitude $A_{n_{+},n_{-}}$
(the $(i^{\pm })$-th line is off-shell). Explicitly, using (\ref{4pt vertices}
) we get (note that the first and second terms on the right-hand side of (
\ref{A+- decomposition}) do not contribute since $p_{2}$ and $p_{1}$ are on
shell) 
\begin{eqnarray}
&& A_{+-}(u,v) =-\frac{R}{g^{2}}\sum\limits_{i^{+}}\frac{u\left( 1\cdot 
\widehat{i}\right) }{uv(1\cdot 2)+u\left( 1\cdot \widehat{i}\right) +v\left(
2\cdot \widehat{i}\right) }\widehat{A}_{n_{+},n_{-}}^{(i^{+})}  \\
&&\hspace{2.2cm} -\frac{R}{g^{2}}\sum\limits_{i^{-}}\frac{v\left( 2\cdot \widehat{i}\right) 
}{uv(1\cdot 2)+u\left( 1\cdot \widehat{i}\right) +v\left( 2\cdot \widehat{i}
\right) }\widehat{A}_{n_{+},n_{-}}^{(i^{-})} +\,\mathcal{R}_{+-}\left( u,v\right) .\notag
\end{eqnarray}
From this we can derive the sequential soft limits, 
\begin{equation}
\lim_{v\rightarrow 0}\lim_{u\rightarrow 0}A_{+-}(u,v)=-\frac{R}{g^{2}}
n_{-}A_{n^{+},n^{-}}+\lim_{v\rightarrow 0}\lim_{u\rightarrow 0}\mathcal{R}
_{+-}\left( u,v\right) ,  \label{uv limit}
\end{equation}
where the amplitude $A_{n^{+},n^{-}}$ corresponds to the original $
(n_{+}+n_{-})$-particle configuration $p_{i}=$ $\widehat{p}_{i}\left(
0,0\right) $. Similarly, in reversed order, 
\begin{equation}
\lim_{u\rightarrow 0}\lim_{u\rightarrow v}A_{+-}(u,v)=-\frac{R}{g^{2}}
n_{+}A_{n^{+},n^{-}}+\lim_{u\rightarrow 0}\lim_{u\rightarrow v}\mathcal{R}
_{+-}\left( u,v\right) .  \label{vu limit}
\end{equation}
To proceed further, we need some information on the double soft limits of
the remainder $\mathcal{R}_{+-}\left( u,v\right) $. First we show, that the
two limits of $\mathcal{R}_{+-}\left( u,v\right) $\ on the right-hand sides
of (\ref{uv limit}) and (\ref{vu limit}) coincide. Indeed, on the one hand we get 
\begin{equation}
\left[ \lim_{v\rightarrow 0},\lim_{u\rightarrow 0}\right] A_{+-}(u,v)=\frac{R
}{g^{2}}\left( n_{+}-n_{-}\right) A_{n_{+},n_{-}}+\left[ \lim_{v\rightarrow
0},\lim_{u\rightarrow 0}\right] \mathcal{R}_{+-}\left( u,v\right).
\end{equation}
On the other hand, we can calculate the left-hand side of the last equation
using the single soft theorems applying them twice. According to (\ref
{general_soft_theorem+}) and (\ref{general_soft_theorem-}),
\begin{eqnarray}
&& \left[ \lim_{v\rightarrow 0},\lim_{u\rightarrow 0}\right] A_{+-}(u,v) = \\
&&\hspace{0.4cm} =g^{-1/2}\left[ -\frac{1}{2}\left( n^{+}-n^{-}-1\right) g^{-1}\left(
\partial _{\overline{w}}g\right) +\partial _{\overline{w}}\right]  
\times\, g^{-1/2}\left[ -\frac{1}{2}\left( n^{-}-n^{+}\right) g^{-1}\left( \partial
_{w}g\right) +\partial _{w}\right] A_{n^{+},n^{-}}\notag \\
&& \hspace{0.5cm}-g^{-1/2}\left[ -\frac{1}{2}\left( n^{-}-n^{+}-1\right) g^{-1}\left(
\partial _{w}g\right) +\partial _{w}\right]  \times\, g^{-1/2}\left[ -\frac{1}{2}\left( n^{+}-n^{-}\right) g^{-1}\left( \partial
_{\overline{w}}g\right) +\partial _{\overline{w}}\right] A_{n^{+},n^{-}}. 
\notag
\end{eqnarray}
After some algebra we arrive at
\begin{equation}
\left[ \lim_{v\rightarrow 0},\lim_{u\rightarrow 0}\right] A_{+-}(u,v)=\frac{R
}{g^{2}}\left( n_{+}-n_{-}\right) A_{n_{+},n_{-}}\,,  \label{uv-vu}
\end{equation}
therefore the commutator of two sequential soft limits probes the curvature
of the target space, as discussed  in detail in Ref. \cite{Cheung:2021yog}. The result (\ref{uv-vu}) is in accord with 
\begin{equation}
\lim_{v\rightarrow 0}\lim_{u\rightarrow 0}\mathcal{R}_{+-}\left( u,v\right)
=\lim_{u\rightarrow 0}\lim_{u\rightarrow v}\mathcal{R}_{+-}\left( u,v\right)
,  \label{limit order}
\end{equation}
as claimed. As we have seen, the sequential double soft limit is simple in the sense
that we can calculate it using the single soft theorems. Let us now proceed
to the \textquotedblleft true\textquotedblright\ double soft limit when both
soft momenta are sent to zero simultaneously with the same rate. That means
we take now $u=v=t$. We then arrive at 
\begin{equation*}
\lim_{t\rightarrow 0}A_{+-}(t,t) =-\frac{R}{g^{2}}\left[
\sum\limits_{i^{+}}\frac{\left( 1\cdot i\right) }{\left( 1\cdot i\right)
+\left( 2\cdot i\right) }+\sum\limits_{i^{-}}\frac{\left( 2\cdot i\right) }{
\left( 1\cdot i\right) +\left( 2\cdot i\right) }\right] A_{n_{+},n_{-}}+ \mathcal{R}_{+-}\left( 0,0\right).
\end{equation*}
According to (\ref{limit order}) we can calculate $\mathcal{R}_{+-}\left(
0,0\right) $ using e.g. the sequence of limits $\lim_{v\rightarrow
0}\lim_{u\rightarrow 0}$ and formula (\ref{uv limit}) 
\begin{eqnarray}
\mathcal{R}_{+-}\left( 0,0\right) &=&\lim_{v\rightarrow 0}\lim_{u\rightarrow
0}A_{+-}(u,v)+\frac{R}{g^{2}}n_{-}A_{n^{+}+n^{-}}  \\
&=&-g^{-1/2}\left[ -\frac{1}{2}\left( n^{+}-n^{-}-1\right) g^{-1}\left(
\partial _{\overline{w}}g\right) +\partial _{\overline{w}}\right]  \notag \\
&&\times\, g^{-1/2}\left[ -\frac{1}{2}\left( n^{-}-n^{+}\right) g^{-1}\left( \partial
_{w}g\right) +\partial _{w}\right] A_{n^{+},n^{-}}+\frac{R}{g^{2}}
n_{-}A_{n^{+},n^{-}}.\nonumber
\end{eqnarray}
After some algebra we obtain
\begin{eqnarray}
\mathcal{R}_{+-}\left( 0,0\right) &=&\left[ -\frac{1}{4g^{3}}\left(
n^{+}-n^{-}\right) ^{2}\partial_w g\partial_{\overline{w}} g+\frac{R}{2g^{2}}\left( n^{+},n^{-}\right) \right. \\
&&\hspace{1.5cm} \left. +\frac{1}{2g^{2}}\left( n^{+}-n^{-}\right) \left( \partial
_{w}g\partial _{\overline{w}}-\partial _{\overline{w}}g\partial _{w}\right) +
\frac{1}{g}\partial _{\overline{w}}\partial _{w}\right] A_{n^{+}+n^{-}} \,.
\notag
\end{eqnarray}
Finally, the double soft theorem takes the form
\begin{align}
\lim_{t\rightarrow 0}A_{+-}(t,t) &=\left[ -\frac{R}{g^{2}}\left(
\sum\limits_{i^{+}}\frac{\left( 1\cdot i\right) }{\left( 1\cdot i\right)
+\left( 2\cdot i\right) }+\sum\limits_{i^{-}}\frac{\left( 2\cdot i\right) }{
\left( 1\cdot i\right) +\left( 2\cdot i\right) }\right)  -\frac{1}{4g^{3}}\left( n^{+}-n^{-}\right) ^{2}\partial
_{w}g\partial _{\overline{w}}g \right. \notag\\
&\hspace{-0.7cm} \left.+\frac{R}{2g^{2}}\left( n^{+}+n^{-}\right)
 +\frac{1}{2g^{2}}\left( n^{+}-n^{-}\right) \left( \partial
_{w}g\partial _{\overline{w}}-\partial _{\overline{w}}g\partial _{w}\right) +
\frac{1}{g}\partial _{\overline{w}}\partial _{w}\right] A_{n^{+},n^{-}}\,.
\end{align}
Similarly, assuming the amplitude 
\begin{equation*}
A_{++}\left( u,v\right) \equiv A(u1^{+},v2^{+},\ldots ,\widehat{i}
(u,v),\ldots )
\end{equation*}
with $n_{+}+2$ positively charged (two of them are soft) and $n_{-}$ negatively charged particles, we get using the same method
\begin{equation*}
\lim_{t\rightarrow 0}A_{++}\left( t,t\right) =-\frac{n_{+}}{g^{3}}\left(
\partial _{\overline{w}}\partial _{\overline{w}}g-3\partial _{\overline{w}
}g\partial _{\overline{w}}g\right) A_{n_{+},n_{-}}+\mathcal{R}_{++}\left(
0,0\right)
\end{equation*}
where the first term on the right-hand side corresponds to the contributions
of the 4pt vertices with two soft particles as above. In this case we get on
one hand
\begin{equation*}
\left[ \lim_{v\rightarrow 0},\lim_{u\rightarrow 0}\right] A_{++}\left(
u,v\right) =\left[ \lim_{v\rightarrow 0},\lim_{u\rightarrow 0}\right] 
\mathcal{R}_{++}\left( u,v\right)
\end{equation*}
while, on the other hand, using (\ref{general_soft_theorem+}), we have immediately 
\begin{equation*}
\left[ \lim_{v\rightarrow 0},\lim_{u\rightarrow 0}\right] A_{++}\left(
u,v\right) =0.
\end{equation*}
Thus again
\begin{equation*}
\lim_{v\rightarrow 0}\lim_{u\rightarrow 0}\mathcal{R}_{++}\left( u,v\right)
=\lim_{u\rightarrow 0}\lim_{u\rightarrow v}\mathcal{R}_{++}\left( u,v\right).
\end{equation*}
For $\mathcal{R}\left( 0,0\right) $ we get therefore, using twice the single
soft theorems, as above,
\begin{eqnarray}
\mathcal{R}_{++}\left( 0,0\right) &=&g^{-1/2}\left[ -\frac{1}{2}\left(
n^{+}+1-n^{-}\right) g^{-1}\left( \partial _{\overline{w}}g\right) +\partial
_{\overline{w}}\right]  \notag \\
&&\times\, g^{-1/2}\left[ -\frac{1}{2}\left( n^{+}-n^{-}\right) g^{-1}\left( \partial
_{\overline{w}}g\right) +\partial _{\overline{w}}\right] A_{n_{+},n_{-}} 
\notag \\
&&+\frac{n_{+}}{g^{3}}\left( \partial _{\overline{w}}\partial _{\overline{w}
}g-3\partial _{\overline{w}}g\partial _{\overline{w}}g\right) A_{n_{+}n_{-}}\,.
\end{eqnarray}
Finally, assembling all the ingredients together
\begin{eqnarray*}
\lim_{t\rightarrow 0}A_{++}\left( t,t\right) &=&g^{-1/2}\left[ -\frac{1}{2}
\left( n^{+}+1-n^{-}\right) g^{-1}\left( \partial _{\overline{w}}g\right)
+\partial _{\overline{w}}\right] \\[1mm]
&&\times\, g^{-1/2}\left[ -\frac{1}{2}\left( n^{+}-n^{-}\right) g^{-1}\left( \partial
_{\overline{w}}g\right) +\partial _{\overline{w}}\right] A_{n_{+},n_{-},}
\end{eqnarray*}
i.e. the double soft limit reduces to two sequential single soft limits. Let us now apply these results to the model with Lagrangian (\ref{L^K_psi}) discussed in Section~\ref{non GB soft}. Fixing $w=\overline{w}=1/2\alpha $
for simplicity, i.e \ for the theory with Lagrangian (\ref{w=1/2a}), we get
double soft theorems for non-Goldstone bosons in the simple form without
derivatives
\begin{eqnarray}
&&\lim_{t\rightarrow 0}A_{+-}(t,t) =\alpha ^{2}K\left( \sum\limits_{i^{+}}
\frac{\left( 1\cdot i\right) }{\left( 1\cdot i\right) +\left( 2\cdot
i\right) }+\sum\limits_{i^{-}}\frac{\left( 2\cdot i\right) }{\left( 1\cdot
i\right) +\left( 2\cdot i\right) }\right) A_{n^{+},n^{-}}   \\
&&\hspace{2cm}+\frac{1}{4}\alpha ^{2}\Big[ (K{+}2)^{2}(2{-}n_{+}{-}n_{-})^{2}{-}K^{2}\left(
n_{+}{-}n_{-}\right) ^{2}  {+} 4\left( n_{+}{+}n_{-}\right) {-}4(K{+}2)\Big]
A_{n^{+},n^{-}}\, ,\notag
\end{eqnarray}
and
\begin{eqnarray}
\lim_{t\rightarrow 0}A_{++}\left( t,t\right) &=&\frac{1}{4}\alpha ^{2}\left(
-K\left( n_{+}-n_{-}+1\right) +(K+2)(2-n_{+}-n_{-}-1\right) )  \notag \\[2mm]
&&\times \left( -K\left( n_{+}-n_{-}\right) +(K+2)(2-n_{+}-n_{-}\right)
A_{n_{+},n_{-}}\,.
\end{eqnarray}
The case of the general vacuum expectation values $w, \,\,\,\overline{w}$ can be easily
recovered by the substitution $\alpha \rightarrow \alpha ^{-K/2}\left( w+
\overline{w}\right) ^{-(K+2)/2}$ (cf. Section \ \ref{non GB soft}). The
Goldstone boson limit $K\rightarrow -2$ gives even simpler results (note
that in this case the amplitudes vanish for $n_{+}\neq n_{-}$)
\begin{equation}
\lim_{t\rightarrow 0}A_{+-}(t,t) =-2\alpha ^{2}\left( \sum\limits_{i^{+}}
\frac{\left( 1\cdot i\right) }{\left( 1\cdot i\right) {+}\left( 2\cdot
i\right) }{+}\sum\limits_{i^{-}}\frac{\left( 2\cdot i\right) }{\left( 1\cdot
i\right) {+}\left( 2\cdot i\right) }\right) A_{n^{+},n^{-}}  {+} \alpha ^{2}\left( n_{+}{+}n_{-}\right) A_{n^{+},n^{-}}
\end{equation}
and trivially
\begin{equation*}
\lim_{t\rightarrow 0}A_{(n_{+}+2)~n_{-}}(t1^{+},t2^{+},\ldots ,\widehat{i}%
(t,t),\ldots )=0\, .
\end{equation*}

\section{Summary and Outlook}\label{sec5}

In this paper, we analyzed in detail the soft behavior of the non-linear sigma models with two-dimensional target spaces in the perturbative regime.
First, we focused on the models describing the dynamics of two Goldstone bosons corresponding to spontaneous symmetry breaking according to the pattern $\mathcal{G}\rightarrow \mathcal{H}$, i.e. when the target space is a two-dimensional coset $\mathcal{G}/\mathcal{H}$. Apparently, in such a case, the violation of the Adler zero might be possible according to the general single-particle soft theorem \cite{Kampf:2019mcd}; nevertheless, the non-linear sigma models with two-dimensional target space appear to be specific. Here, the soft theorem in its general form puts strong constraints on the amplitudes. Namely, it forbids the odd-point amplitudes to be nonzero. Therefore, the Adler zero cannot be violated and the $U(1)$ symmetry of $S-$matrix is always present, even when not manifest at the Lagrangian level. Since the Adler zero together with power counting and factorization defines the theory uniquely in terms of the four-point seed amplitude, this result can be used for classification of the possible sigma models with two Goldstone bosons. Actually, we have only two possible realizations (up to an admissible field
redefinition), namely the $\mathbb{C}P^{1}$ model with target space $O(3)/O(2)\approx $ $S^{2}$ and its non-compact variant with target space $O(1,2)/O(2)\approx H^{2}$.

As a next step, we studied the soft theorem for the case when the sigma model describes dynamics of non-Goldstone particles, i.e., when the target space is more general and possesses fewer symmetries. As a first step, we introduced a special example of the one-parametric class of non-linear sigma models. These generalize the Goldstone boson case. In general, the Goldstone bosons couple to the conserved currents which are associated with the Killing vectors of the target space manifold. Namely this feature is responsible for the validity of single-particle soft theorems (the Adler zero). In our class of models, the target space possesses one Killing vector, and on top of this also two additional conformal Killing vectors forming together the $U(1,1)$ algebra. The particles then couple to the currents corresponding to the Killing vector of the shift symmetry and to one of the conformal Killing vectors. This allows us to prove a new type of single-particle soft theorem for our model, which substantially differs from the usual one for Goldstone bosons: the soft limit of the $n-$point amplitude is still expressed in terms of the $(n-1)$-pt amplitudes, but the multiplicative soft factor depends on the number of particles indicating the factorial growth of the amplitudes with the particle multiplicity. The class of these models is broad enough to encompass either the $O(1,2)/O(2)$ model, which appears as a special limit for which the conformal Killing vector becomes the ordinary Killing one, or a special Lie-algebraic sigma model (\ref{Lie-algebraic model}).

Then we discussed possible generalizations. First, we concentrated on slightly more general cases, which partially share the features of the above special class of sigma models. Namely we supposed the presence of (conformal) Killing vectors and their currents to which some linear combinations of the one-particle states couple. In such a case, the soft sum rules can be proved, which express soft limits of linear combinations of amplitudes with different flavors in terms of the lower point amplitudes.

Finally we assumed the general case without (conformal) Killing vectors. Such models still allow to formulate a single-particle soft theorem, the price to pay here is the necessity to know the $S-$ matrix as a function of the classical vacuum, i.e. point in the target space around which we perform the weak field expansion. This piece of information is equivalent to the knowledge of the full Lagrangian of the theory and therefore, the identification of the theory by the soft theorem avoids its reconstruction by soft recursion. We presented an elementary proof of such a soft theorem in the case of general two-dimensional K\"ahlerian target space and relate it to the geometric soft theorem discussed in \cite{Cheung:2021yog}. We also discussed the double soft limit in the general case, the formulation of which again requires the knowledge of the dependence of the amplitudes on the classical vacuum. Nevertheless, provided the target space is a subject of appropriate symmetries, both these theorems simplify and the knowledge of the derivatives of the amplitudes with respect to the classical vacuum is not necessary. As a byproduct, we can then derive the action of the (conformal) Killing vectors on the amplitudes as functions of the classical vacuum. We provided an illustration of this mechanism using the above mentioned class of sigma models with (conformal) Killing vectors on the target space. 

\subsection*{Outlook}

Our paper shows a particular set of non-Goldstone models with special soft theorems. At the moment, there is no systematic way how to search and analyze such models and provide a classification of interesting non-linear sigma models from an amplitudes perspective, especially for the $n$ scalar case. The general soft theorem as discussed in \cite{Cheung:2021yog} contains derivatives of amplitudes on the right-hand side. Based on our experience with the two-scalar case, the criterion for the existence of the special case is the absence of these derivatives, ie. the soft theorem only contains amplitudes (not their derivatives). This would be likely a consequence of a certain underlying symmetry of the model. Finally, it would be interesting to explore the connection of these special models (including the ones found in this paper) to other amplitudes methods, especially the color-kinematics duality and CHY formalism.

\section*{Acknowledgments}

M.S. is grateful to O. Gamayun and A. Losev for useful discussions. This work is supported in part by DOE grants DE-SC0011842 and DE-SC0009999, GA\-\v{C}R 24-11722S, MEYS LUAUS23126, OP JAK CZ.02.01.01/00/22\_008/0004632, the Simons Foundation Targeted Grant 920184 to the Fine Theoretical Physics Institute, and the funds of the University of California.

A part of this work has been carried out at the Institute of Particle and Nuclear Physics, Charles University, during the stay of M.S. in Prague in 2023 in the framework of the Fulbright Distinguished Scholars Program. The Fulbright Foundation support is highly appreciated. Sincere thanks for warm hospitality go to the Institute of Particle and Nuclear Physics in Prague and to Kate\v{r}ina Kloubov\'a, a program officer at the Czech Fulbright Commission.

\appendix

\section{Derivation of the general soft theorem; see Secs. \ref{nngm}, \ref{43}}\label{appA}

The relevant momentum space Feynman rules for the insertions of $\widetilde{J%
}\left( p\right) $
\begin{equation*}
J=\partial z+g^{-3/2}z\partial z\partial _{w}g+g^{-3/2}\overline{z}\partial
z\partial _{\overline{w}}g+\ldots
\end{equation*}%
 are given by (\ref{currents general}) which reads (all the momenta are treated as outgoing)%
\begin{eqnarray}
\partial z &\rightarrow &V_{2}\left( p,-p\right) =-\mathrm{i}p \,, \notag \\[1mm]
g^{-3/2}\partial _{w}g~z\partial z &\rightarrow &V_{3}(p,k,-k-p)=-\mathrm{i}%
p~g^{-3/2}\partial _{w}g \,, \notag \\[1mm]
g^{-3/2}\partial _{\overline{w}}g~\overline{z}\partial z &\rightarrow
&V_{3}(p,k,\overline{-k-p})=\mathrm{i}k~g^{-3/2}\partial _{\overline{w}}g\,.
\label{current Feynman rules}
\end{eqnarray}%
Then%
\begin{equation}
\langle 0|J\left( 0\right) |p^{+}\rangle =-\mathrm{i}p,~~~\langle 0|%
\overline{J}\left( 0\right) |p^{-}\rangle =-\mathrm{i}p
\end{equation}%
and thus for the out state $|n^{+},n^{-}\rangle =|1^{+},\ldots
,n^{+},1^{-},\ldots ,n^{-}\rangle $ we get 
\begin{equation}
\left\langle n^{+},n^{-}|{J}\left( 0\right) |0\right\rangle =-%
\mathrm{i}pg^{1/2}\frac{\mathrm{i}}{p^{2}}\mathrm{i}A_{n^{+}+n^{-}+p^{+}}+{\cal{R}}
\end{equation}%
where%
\begin{equation}
\mathrm{i}A_{n^{+}+1,n^{-}}=\left\langle p^{+},n^{+},n^{-}|0\right\rangle .
\end{equation}%
Thus, the balance equation of the current gives  
\begin{equation}
\lim_{p\rightarrow 0}A_{n^{+}+1,n^{-}}=\lim_{p\rightarrow 0}\mathrm{i}p\cdot
{\cal{R}}+g^{-1/2}\int \mathrm{d}^{D}x\left\langle n^{+},n^{-}|\partial _{\overline{w%
}}\mathcal{L}_{w}^{int}\left( x\right) |0\right\rangle \,.
\end{equation}%
The relevant part of ${\cal{R}}$ comes from insertions of the current $J$ into
external lines. The only nonvanishing contribution to $\mathrm{i}p\cdot{\cal{R}}$
in the soft limit stems from the terms in $J$ which are quadratic in the
fields. Using (\ref{current Feynman rules}), we get%
\begin{eqnarray*}
{\cal{R}} &=&g^{-3/2}\partial _{w}g\sum\limits_{i=1}^{n^{-}}\left( -\mathrm{i}%
p\right) \frac{\mathrm{i}}{(p+k_{i})^{2}}\mathrm{i}A_{n^{+},n^{-}}(1^{+}%
\ldots ,\left( p+k_{i}\right) ^{+},\ldots n^{-}) \\
&&+g^{-3/2}\partial _{\overline{w}}g\sum\limits_{i=1}^{n^{+}}-\mathrm{i}%
\left( p+k_{i}\right) \frac{\mathrm{i}}{(p+k_{i})^{2}}\mathrm{i}%
A_{n^{+},n^{-}}(1^{+}\ldots ,\left( p+k_{i}\right) ^{+},\ldots n^{-}) \\
&&+g^{-3/2}\partial _{\overline{w}}g\sum\limits_{i=1}^{n^{-}}\mathrm{i}k_{i}%
\frac{\mathrm{i}}{(p+k_{i})^{2}}\mathrm{i}A_{n^{+},n^{-}}(1^{+}\ldots
,\left( p+k_{i}\right) ^{-},\ldots n^{-})+O\left( p\right)
\end{eqnarray*}%
where the individual rows correspond to the insertion of $g^{-3/2}\partial
_{w}g~z\partial z$ into the external $z$ line, insertion of $%
g^{-3/2}\partial _{\overline{w}}g~\overline{z}\partial z$ into external $%
\overline{z}$ line and insertion of $g^{-3/2}\partial _{\overline{w}}g~%
\overline{z}\partial z$ into external $z$ line respectively. Therefore 
\begin{equation}
\lim_{p\rightarrow 0}\mathrm{i}p\cdot{\cal{R}}=-\frac{1}{2}\left(
n^{+}-n^{-}\right) g^{-1}\left( \partial _{\overline{w}}g\right)
A_{n^{+},n^{-}}(1^{+}\ldots ,n^{-})\,.
\end{equation}%
We also need integrated matrix element of the operator $\partial _{\overline{%
w}}\mathcal{L}_{w}^{int}\left( x\right) $. Using the Dirac picture in the
intermediate steps we obtain 
\begin{eqnarray*}
\int \mathrm{d}^{D}x\left\langle n^{+},n^{-}|\partial _{\overline{w}}%
\mathcal{L}_{w}^{int}\left( x\right) |0\right\rangle &=&\int \mathrm{d}%
^{D}x\left\langle n^{+},n^{-}\left|T\partial _{\overline{w}}\mathcal{L}%
_{w}^{int}\left( x\right) _{D}\exp \left( \mathrm{i}\int \mathrm{d}^{D}y%
\mathcal{L}_{w}^{int}\left( y\right) _{D}\right) \right|0\right\rangle \\[2mm]
&=&-\mathrm{i}\partial _{\overline{w}}\int \mathrm{d}^{D}x\left\langle
n^{+},n^{-}\left|T\partial _{\overline{w}}\exp \left( \int \mathrm{d}^{D}y%
\mathcal{L}_{w}^{int}\left( y\right) _{D}\right) \right|0\right\rangle \\[2mm]
&=&-\mathrm{i}\partial _{\overline{w}}\left\langle
n^{+},n^{-}|0\right\rangle =\partial _{\overline{w}}A_{n^{+},n^{-}}(1^{+}%
\ldots ,n^{-}).
\end{eqnarray*}%
As a result we get the soft theorem in the form (\ref{general_soft_theorem+}%
).

\section{Example of the soft theorems for 5pt amplitudes}\label{appB}

As a particular example, let us examine the Lagrangian%
\begin{equation}
\mathcal{L}=\alpha ^{K}\left( z+\overline{z}\right) ^{K}\partial z\cdot
\partial \overline{z}\,.
\end{equation}%
Then%
\begin{equation}
\mathcal{L}_{w}=\alpha ^{K}\left( z+\overline{z}+w+\overline{w}\right)
^{K}\partial z\cdot \partial \overline{z}
\end{equation}%
or after normalization of the kinetic term by field redefinition $%
z\rightarrow z\alpha ^{-K/2}\left( w+\overline{w}\right) ^{-K/2}$ we get%
\begin{eqnarray}
\mathcal{L}_{w} &=&\alpha ^{K}\left( w+\overline{w}+\frac{1}{\alpha
^{K/2}\left( w+\overline{w}\right) ^{K/2}}\left( z+\overline{z}\right)
\right) ^{K}\frac{\partial z\cdot \partial \overline{z}}{\alpha ^{K}\left( w+%
\overline{w}\right) ^{K}}  \notag \\
&=&\left( 1+\frac{1}{\alpha ^{K/2}\left( w+\overline{w}\right) ^{1+K/2}}%
\left( z+\overline{z}\right) \right) ^{K}\partial z\cdot \partial \overline{z%
}
\end{eqnarray}%
and in the previous formulae $g=\alpha ^{K}\left( w+\overline{w}\right) ^{K}$. 
Expanding in $z,\overline{z}$ we arrive at
\begin{equation}
\mathcal{L}_{w}=\partial z\cdot \partial \overline{z}\left( 1+\frac{K}{%
\alpha ^{K/2}\left( w+\overline{w}\right) ^{1+K/2}}\left( z+\overline{z}%
\right) +\ldots \right).
\end{equation}%
To simplify the calculations, let us make further change of variables to
avoid the cubic vertices. The transformation reads%
\begin{eqnarray}
z &\rightarrow &z-z^{2}\frac{K}{2\alpha ^{K/2}\left( w+\overline{w}\right)
^{1+K/2}} \,, \notag \\[1mm]
z &\rightarrow &\overline{z}-\overline{z}^{2}\frac{K}{2\alpha ^{K/2}\left( w+%
\overline{w}\right) ^{1+K/2}}\,,
\end{eqnarray}%
and using these new variables, the Lagrangian takes the form
\begin{eqnarray}
\mathcal{L}_{w} &=&\partial z\cdot \partial \overline{z}-\frac{K}{2\alpha
^{K}\left( w+\overline{w}\right) ^{K+2}}\left[ \left( 1+2K\right) \left(
z^{2}+\overline{z}^{2}\right) +2z\overline{z}\right] \partial z\cdot
\partial \overline{z}  \notag \\[1mm]
&&-\frac{1}{6}\frac{K}{\left( w+\overline{w}\right) ^{3\left( 1+K/2\right)
}\alpha ^{3K/2}}\left( z+\overline{z}\right) \partial z\cdot \partial 
\overline{z}  \notag \\[1mm]
&&\times \left[ \left( K-2\right) \left( 1+2K\right) \left( z^{2}+\overline{z%
}^{2}\right) -2\left( K^{2}+2\right) z\overline{z}\right] +O\left(
z^{6}\right) .
\end{eqnarray}%
There is just one nontrivial 4pt amplitude, namely,%
\begin{equation}
A_{2,2}\left( 1^{+},2^{+},3^{-},4^{-}\right) =\frac{K}{\alpha ^{K}\left( w+%
\overline{w}\right) ^{K+2}}\left( 1+2\right) \cdot \left( 3+4\right),
\end{equation}%
and two 5pt amplitudes%
\begin{eqnarray}
A_{3,2}\left( p^{+},1^{+},2^{+},3^{-},4^{-}\right) =-\frac{K\left( K+2\right) }{\left( w+\overline{w}\right) ^{3\left(
1+K/2\right) }\alpha ^{3K/2}}\left( p+1+2\right) \cdot \left( 3+4\right), 
\end{eqnarray}%
and%
\begin{equation}
A_{2,3}\left( 1^{+},2^{+},3^{-},4^{-},p^{-}\right) =-\frac{K\left( K+2\right) 
}{\left( w+\overline{w}\right) ^{3\left( 1+K/2\right) }\alpha ^{3K/2}}\left(
1+2\right) \cdot \left( 3+4+p\right).
\end{equation}%
Then e.g.%
\begin{eqnarray}
\lim_{p\rightarrow 0}A_{3,2}\left( p^{+},1^{+},2^{+},3^{-},4^{-}\right) &=&-%
\frac{K\left( K+2\right) }{\left( w+\overline{w}\right) ^{3\left(
1+K/2\right) }\alpha ^{3K/2}}\left( 1+2\right) \cdot \left( 3+4\right) 
\notag \\[1mm]
&=&-\frac{K+2}{\alpha ^{K/2}\left( w+\overline{w}\right) ^{K/2+1}}%
A_{2,2}\left( 1^{+},2^{+},3^{-},4^{-}\right). \notag \\
\label{5pt_soft_limit}
\end{eqnarray}%
In this case, the relevant soft factor on the right-hand side of (\ref{soft_theorem}) reads
\begin{eqnarray}
\frac{1}{2}{\alpha ^{-K/2}\left( w+\overline{w}\right) ^{-1-K/2}} &&\left( -K\left( n^{+}-n^{-}\right) +\left( K+2\right)
(2-n^{+}-n^{-})\right) \big\vert_{n^+=n^-=2} \notag \\[2mm]
&=&-\frac{K+2}{\alpha ^{K/2}\left( w+\overline{w}\right) ^{1+K/2}}
\end{eqnarray}%
in accord with (\ref{5pt_soft_limit}). On the other hand,  
 the right-hand side of (\ref{soft_theorem}) is
\begin{eqnarray}
&&g^{-1/2}\left[ -\frac{1}{2}\left( n^{+}-n^{-}\right) g^{-1}\left( \partial
_{\overline{w}}g\right) +\partial _{\overline{w}}\right]
A_{n^{+},n^{-}}(n^{+}=n^{-}=2)  \notag \\[1mm]
&=&\alpha ^{-K/2}\left( w+\overline{w}\right) ^{-K/2}\partial _{\overline{w}}
\frac{K}{\alpha ^{K}\left( w+\overline{w}\right) ^{K+2}}\left( 1+2\right)
\cdot \left( 3+4\right)  \notag \\[1mm]
&=&-\frac{K\left( K+2\right) }{\left( w+\overline{w}\right) ^{3\left(
1+K/2\right) }\alpha ^{3K/2}}\left( 1+2\right) \cdot \left( 3+4\right),
\end{eqnarray}
as expected from (\ref{5pt_soft_limit}).

\newpage

\bibliography{NLSM}
\bibliographystyle{JHEP}

\end{document}